\documentclass[longauth,doublecolumn]{aa} 
\usepackage{graphicx}
\usepackage{txfonts}
\usepackage{savesym}
\savesymbol{rotate}
\usepackage{deluxetable} 
\restoresymbol{DTBL}{rotate}

\usepackage[english]{babel}
\usepackage{times,epsfig}
\usepackage{longtable}
\usepackage{url}

\usepackage{mathrsfs}
\usepackage[normalem]{ulem}

\usepackage{soul}
\usepackage{ulem}

\usepackage{hyperref}
\hypersetup{
    colorlinks = true,
    linkcolor = blue,
    anchorcolor = blue,
    citecolor = blue,
    filecolor = blue,
    urlcolor = blue
    }

\usepackage{placeins}

\newcommand{\orcid}[1]{\href{https://orcid.org/#1}{\includegraphics[width=10pt]{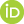}}}

\usepackage{multirow}
\usepackage{booktabs}
\usepackage{array}
\usepackage{amsmath}

\def\apjl{ApJL}%
\def\apjs{ApJS}%
\def\ao{Appl.~Opt.}%
\def\aap{A\&A}%

\DeclareMathAlphabet{\mathsc}{OT1}{cmr}{m}{sc}
\def\testbx{bx}%
\DeclareRobustCommand{\ion}[2]{%
\relax\ifmmode
\ifx\testbx\f@series
{\mathbf{#1\,\mathsc{#2}}}\else
{\mathrm{#1\,\mathsc{#2}}}\fi
\else\textup{#1\,{\mdseries\textsc{#2}}}%
\fi}

\newcommand{\dmb} {\mbox{$\Delta m_{15}(B)$}}
\newcommand{\sbv} {\mbox{$s_{BV}$}}

\newcommand{\ha} {\mbox{H$\alpha$}}
\newcommand{\hb} {\mbox{H$\beta$}}

\newcommand{\Oi} {\ion{O}{i}}

\newcommand{\Nai} {\ion{Na}{i}}

\newcommand{\Siii} {\ion{Si}{ii}}

\newcommand{\Siiii} {\ion{Si}{iii}}

\newcommand{\Caii} {\ion{Ca}{ii}}

\newcommand{\Tiii} {\ion{Ti}{ii}}

\newcommand{\Feii} {\ion{Fe}{ii}}

\newcommand{\sn}{ASASSN-20jq}
\newcommand{\sntns}{SN~2020qxp}
\newcommand{\host}{NGC\,5002}

\newcommand{\PeakEpoch}{JD~2,459,085.8}
\newcommand{\snoopy}{\textsc{SNooPy}}

\newcommand{\bv}{\mbox{$B\!-\!V$}}
\newcommand{\ebv}{\mbox{$E(B-V)$}}

\newcommand{\msun}{\mbox{M$_{\odot}$}}
\newcommand{\msol}{\mbox{M$_{\odot}$}}

\newcommand{\kms}{\mbox{$\rm{\,km\,s^{-1}}$}}

\newcommand{\nickel}{\mbox{$^{56}$Ni}}
\newcommand{\cobalt}{\mbox{$^{56}$Co}}
\newcommand{\iron}{\mbox{$^{56}$Fe}}

\newcommand{\ld}{\mbox{$\lambda$}}
\newcommand{\ldld}{\mbox{$\lambda\lambda$}}

\authorrunning{Bose, Stritzinger et al.}
\titlerunning{The underluminous type-Ia SN \sn.}
\begin{document} 

\title{Expanding the parameter space of  2002es-like type~Ia supernovae: on the underluminous \sn\ / \sntns}

\author{
Subhash Bose\inst{\ref{inst:au},\ref{inst:osu},\ref{inst:ccapp}} \orcid{0000-0003-3529-3854}
\and Maximilian D. Stritzinger\inst{\ref{inst:au}}\orcid{0000-0002-5571-1833}
\and Chris Ashall\inst{\ref{inst:huusa}}\orcid{0000-0002-5221-7557}
\and Eddie Baron\inst{\ref{inst:ptusa},\ref{inst:hsde}}\orcid{0000-0001-5393-1608}
\and Peter Hoeflich\inst{\ref{inst:fsuusa}}\orcid{0000-0002-4338-6586}
\and L. Galbany\inst{\ref{inst:icespan},\ref{inst:ieecspain}}\orcid{0000-0002-1296-6887}
\and W. B. Hoogendam\inst{\ref{inst:huusa}}\orcid{0000-0003-3953-9532}
\and E. A. M. Jensen\inst{\ref{inst:au}}\orcid{0000-0003-3197-3430}
\and C. S. Kochanek\inst{\ref{inst:osu},\ref{inst:ccapp}}\orcid{0000-0001-6017-2961}
\and R. S. Post\inst{\ref{inst:pousa}}\orcid{0000-0003-3244-0337}
\and A. Reguitti\inst{\ref{inst:inafmerate},\ref{inst:inafpadova}}\orcid{0000-0003-4254-2724}
\and N. Elias-Rosa\inst{\ref{inst:inafpadova},\ref{inst:icespan}}\orcid{0000-0002-1381-9125}
\and K. Z. Stanek\inst{\ref{inst:osu},\ref{inst:ccapp}}
\and Peter Lundqvist\inst{\ref{inst:suse}}
\and Katie Auchettl\inst{\ref{inst:muau},\ref{inst:ucscus}}\orcid{0000-0002-4449-915}
\and Alejandro Clocchiatti\inst{\ref{inst:puccl}}\orcid{0000-0003-3068-4258}
\and A. Fiore\inst{\ref{inst:iftpde},\ref{inst:inafpadova}}\orcid{0000-0002-0403-3331}
\and Claudia P. Guti\'errez\inst{\ref{inst:ieecspain2},\ref{inst:icespan}}\orcid{0000-0003-2375-2064}
\and Jason T. Hinkle\inst{\ref{inst:huusa}}\orcid{0000-0001-9668-2920}
\and Mark E. Huber\inst{\ref{inst:huusa}}\orcid{0000-0003-1059-9603}
\and T. de Jaeger \inst{\ref{inst:sufr}} \orcid{0000-0001-6069-1139}
\and Andrea Pastorello\inst{\ref{inst:inafpadova}}\orcid{0000-0002-7259-4624}
\and Anna V. Payne\inst{\ref{inst:huusa}}\orcid{0000-0003-3490-3243}
\and Mark Phillips\inst{\ref{inst:LCO}}\orcid{0000-0003-2734-0796}
\and Benjamin J. Shappee\inst{\ref{inst:huusa}}\orcid{0000-0003-4631-1149}
\and Michael A. Tucker\inst{\ref{inst:ccapp},\ref{inst:osu}}\orcid{0000-0002-2471-8442}
}
\institute{
Department of Physics and Astronomy, Aarhus University, Ny Munkegade 120, DK-8000 Aarhus C, Denmark (\email{email@subhashbose.com, max@phys.au.dk})\label{inst:au}
\and 
Department of Astronomy, The Ohio State University, 140 W. 18th Avenue, Columbus, OH 43210, USA\label{inst:osu}
\and 
Center for Cosmology and AstroParticle Physics (CCAPP), The Ohio State University, 191 W. Woodruff Avenue, Columbus, OH 43210, USA\label{inst:ccapp}
\and 
Institute for Astronomy, University of Hawai\`{}i at Manoa, 2680 Woodlawn Drive, Honolulu, HI 96822, USA\label{inst:huusa}
\and 
Planetary Science Institute, 1700 E Fort Lowell Rd., Ste 106, Tucson, AZ 85719 USA\label{inst:ptusa}
\and
Hamburger Sternwarte, Gojensbergweg 112, 21029 Hamburg, Germany\label{inst:hsde}
\and 
Department of Physics, Florida State University,
77 Chieftain Way, Tallahassee, FL, 32306, USA\label{inst:fsuusa}
\and 
Institute of Space Sciences (ICE, CSIC), Campus UAB, Carrer de Can Magrans, s/n, E-08193 Barcelona, Spain\label{inst:icespan}
\and
Institut d’Estudis Espacials de Catalunya (IEEC), E-08034 Barcelona, Spain\label{inst:ieecspain}
\and
Post Astronomy, Lexington, MA, USA\label{inst:pousa}
\and INAF – Osservatorio Astronomico di Brera, Via E. Bianchi 46, I-23807 Merate (LC), Italy\label{inst:inafmerate}
\and
INAF – Osservatorio Astronomico di Padova, Vicolo dell’Osservatorio 5, I-35122 Padova, Italy\label{inst:inafpadova}
\and Department of Astronomy and The Oskar Klein Centre, AlbaNova University Center, Stockholm University, 106 91 Stockholm, Sweden\label{inst:suse}
\and School of Physics, The University of Melbourne, VIC 3010, Australia\label{inst:muau}
\and Department of Astronomy and Astrophysics, University of California, Santa Cruz, CA 95064, USA\label{inst:ucscus}
\and Pontificia Universidad Catolica de Chile\label{inst:puccl}
\and Institut f\"ur Theoretische Physik, Goethe Universit\"at, Max-von-Laue-Str. 1, 60438 Frankfurt am Main, Germany\label{inst:iftpde}
\and Institut d'Estudis Espacials de Catalunya (IEEC), Edifici RDIT, Campus UPC, 08860 Castelldefels (Barcelona), Spain\label{inst:ieecspain2}
\and Sorbonne Université, CNRS/IN2P3, LPNHE, F-75005, Paris, France\label{inst:sufr}
\and Carnegie Observatories, Las Campanas Observatory, Casilla 601, La Serena, Chile\label{inst:LCO}
}

\date{Received 06 January, 2025 \ Accepted  22 April, 2025.}

\abstract
{We present optical photometric and spectroscopic observations of the peculiar Type~Ia supernovae (SNe~Ia) \sn\ / \sntns. It is a low-luminosity object, having peak absolute magnitude of $M_B=-17.1\pm0.5$\,mag, while the post-peak light-curve decline rate of $\dmb=1.35\pm0.09$\,mag and color-stretch parameter of $\sbv\gtrapprox0.82$ is similar to normal luminosity SNe~Ia. That makes it a prevalent outlier in both the SN~Ia luminosity-width and the luminosity-color-stretch relations. The analysis of the early light curves indicates a possible ``bump''  during the first $\approx1.4$ days of explosion. \sn\ synthesized a low radioactive \nickel\ mass of $0.09\pm0.01$\msun. The near-maximum light spectra of the supernova show strong \Siii\ absorption lines, indicating a cooler photosphere than normal SNe~Ia, however, it lacks \Tiii\ absorption lines. Additionally, it shows unusually strong absorption features of  \Oi\ \ld7773 and the \Caii\ near-infrared triplet. Nebular spectra of \sn\ show a remarkably strong but narrow forbidden [\Caii] \ldld7291, 7324 doublet emission that is not seen in SNe~Ia except for a handful of  Type~Iax events. There is also marginal detection of the [\Oi] \ldld6300, 6364 doublet emission in nebular spectra, which is extremely rare. Both the [\Caii] and [\Oi] lines are redshifted by roughly $2000$\kms.  \sn\ also exhibits a strong [\Feii] \ld7155 emission line with a tilted-top line profile, which is identical to the [\Feii] \ld16433 line profile.
The asymmetric [\Feii] line profiles, along with the
  redshifted [\Caii] and [\Oi] emission lines, suggest a high central
  density white dwarf progenitor that underwent an off-center
  delayed-detonation explosion mechanism, synthesizing roughly equal
  amounts of $^{56}$Ni during the deflagration and detonation burning
  phases. The equal production of $^{56}$Ni in both burning
  phases distinguishes \sn\ from normal bright and subluminous SNe~Ia. Assuming this scenario, we simultaneously modeled the optical and near-infrared nebular spectra, achieving good agreement with the observations.
The light curve and spectroscopic features of \sn\ do not align with any single sub-class of SNe~Ia. However, the significant deviation from the luminosity versus light-curve shape relations along with several light-curve and spectroscopic features, shows similarities to some 2002es-like objects. Therefore, we add \sn\ as an extreme candidate within the broad and heterogeneous parameter space of 2002es-like SNe~Ia.}

\keywords{supernovae: general $-$ supernovae: individual: (\sn, \sntns) $-$ galaxies: individual: \host\ }  

\maketitle 

\section{Introduction} 
\label{sec:intro}

 Type~Ia supernovae (SNe~Ia) are quintessential distance indicators used to map the expansion history of the Universe \citep{Hubble1926}.  SNe~Ia arise from the thermonuclear disruption of one or more carbon-oxygen white dwarf (WD) stars in a binary system   \citep[for contemporary reviews see][and references therein]{Maoz2014,Liu2023}.   As the WD approaches the Chandrasekhar limit a thermonuclear runaway ensues leading to the disruption of the WD, and in the process significant amounts of radioactive $^{56}$Ni are synthesized, which feeds the emission of the bright optical transients which typically reach peak absolute $B$-band magnitudes of $M_B \gtrsim -19\pm1$  \citep[e.g.,][]{1996AJ....112.2391H}. Given their high intrinsic peak brightness and apparent uniformity, SNe~Ia were considered to be an ideal luminosity-distance indicator.
However, with the discovery of the subluminous SNe~1991bg \citep{Filippenko1992,Leibundgut1993} and the overluminous SN~1991T \citep{Filippenko1992,Phillips1992}, it became clear SNe~Ia exhibit considerable diversity among key observational parameters including both peak luminosities and colors \citep[e.g.,][]{Hamuy1994}.

Fortunately, along with expanded samples of SNe~Ia \citep[e.g.,][]{Hamuy1996,Riess1999},  empirical relations were teased out of the  data \citep{1977SvA....21..675P,Phillips1993,Tripp1998},\footnote{Expanding upon earlier work of \citet{1977SvA....21..675P}, \citet{Phillips1993} introduced the luminosity decline-rate parameter $\Delta m_{15}(B)$. This parameter quantifies the change in the brightness of an SN~Ia  as measured from the  $B$-band light curve between the peak and 15 days later.   \citeauthor{Phillips1993} demonstrated $\Delta m_{15}(B)$ correlates with the peak luminosity in the sense that bright SNe~Ia exhibit more slowly declining light curves.} which when leveraged properly, enabled the reduction in the scatter of the SNe~Ia's peak luminosities  to only $ \sim0.1 $\,mag \citep[e.g.][]{1996AJ....112.2391H,Phillips1999,Goldhaber2001}. Since the mid-1990s, SN~Ia samples have expanded \citep[e.g.,][]{Jha2006,Hicken2009,Ganeshalingam2010,Stritzinger2011,Krisciunas2017,Sako2018,2022PASP..134l4502T,2024MNRAS.tmp.2445D}, calibration techniques have improved  \citep[e.g.][]{Guy2007,Jha2007,Mandel2011,Burns2014,Mandel2017,Burns2018}, and  now SNe~Ia  form the bedrock of modern cosmological investigations aimed to precisely measure the Hubble constant \citep[e.g.,][]{Hamuy1995,Freedman2001,Freedman2009,Riess2016} and to quantify the accelerated expansion of the Universe \citep[e.g.,][]{1998AJ....116.1009R,Schmidt1998,Ganeshalingam2013,Betoule2014,Scolnic2018,Burns2018,Riess2021,1999ApJ...517..565P}.

\begin{figure}[!t]
\centering
\includegraphics[width=1\linewidth]{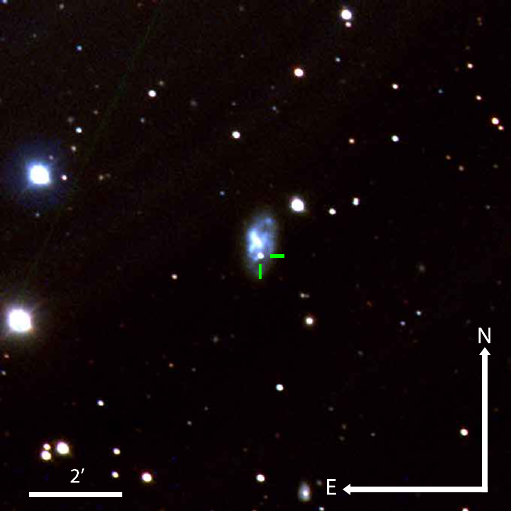}
\caption{ Composite \textit{Bgr}-band ($11\arcmin \times 11\arcmin$) image of \sn\ (green marker) located within \host\ and observed from the Post observatory 15 days prior to $B$-band maximum.}
\label{fig:host_image}
\end{figure}

In recent years, a growing number of additional SN~Ia subtypes and subclasses have been identified in the literature, populated by objects that deviate from the established luminosity-width and color relations  \citep[see reviews by][and references therein]{2017hsn..book..317T,2019NatAs...3..706J}. Mapping the full diversity of peculiar SNe~Ia  is needed to ensure that cosmological parameter estimates based on current and future SNe~Ia samples are not biased by objects departing from the luminosity decline-rate and luminosity color relations, and provides interesting new avenues to elucidate the origins of SNe~Ia progenitor systems and insights on how they explode.   

In this paper, we present detailed optical observations of the type~Ia SN \sn\ (IAU designation SN~2020qxp) discovered soon after the explosion (see Sect.~\ref{sec:discovery}) by the All-Sky Automated Survey for SuperNovae \citep[ASAS-SN;][]{2014ApJ...788...48S,2017PASP..129j4502K}, and initially classified as a transitional SNe~Ia \citep{Stritzinger2020}. 
Our analysis shows  \sn\ is an underluminous SN~Ia with a broader than expected light curve shape, prevalent spectral features associated with intermediate mass elements, and exhibits an early excess of emission within the first days of explosion. These attributes are reminiscent of objects linked to the loosely defined  2002es-like SNe~Ia subclass \citep{Ganeshalingam2012}.
We also build upon the work of \citet{Hoeflich2021}, who modeled the late-phase near-infrared spectrum of \sn\ using a Chandrasekhar-mass ($M_{Ch}$)  carbon-oxygen (C+O) white dwarf (WD)
with a relatively high-central density which is disrupted following an off-center deflagration-to-detonation transition (DDT). In this study, we extend their model to include optical wavelengths as well. After including the effects of macroscopic mixing, the model is found to be in good agreement with both the optical and near-infrared nebular spectra.
 
In Sect.~\ref{sec:discovery}, we discuss the discovery, first detection, and last non-detection of \sn\ as well as the distance to the host galaxy. Sect.~\ref{sec:obsv} provides a description of data collection. The analysis of photometric and spectroscopic data is presented in Sects.~\ref{sec:photometry} and \ref{sec:spectroscopy}, respectively. 
In Sect.~\ref{sec:host}, we discuss the properties of the host galaxy, while Sect.~\ref{sec:discussion} presents the results, compares them with those of other supernovae, and analyzes the nebular spectra using the off-center DDT model.  Finally, we summarize our findings in Sect.~\ref{sec:summary}.

\begin{figure*}[!ht]
	\centering
	\includegraphics[width=0.90\linewidth]{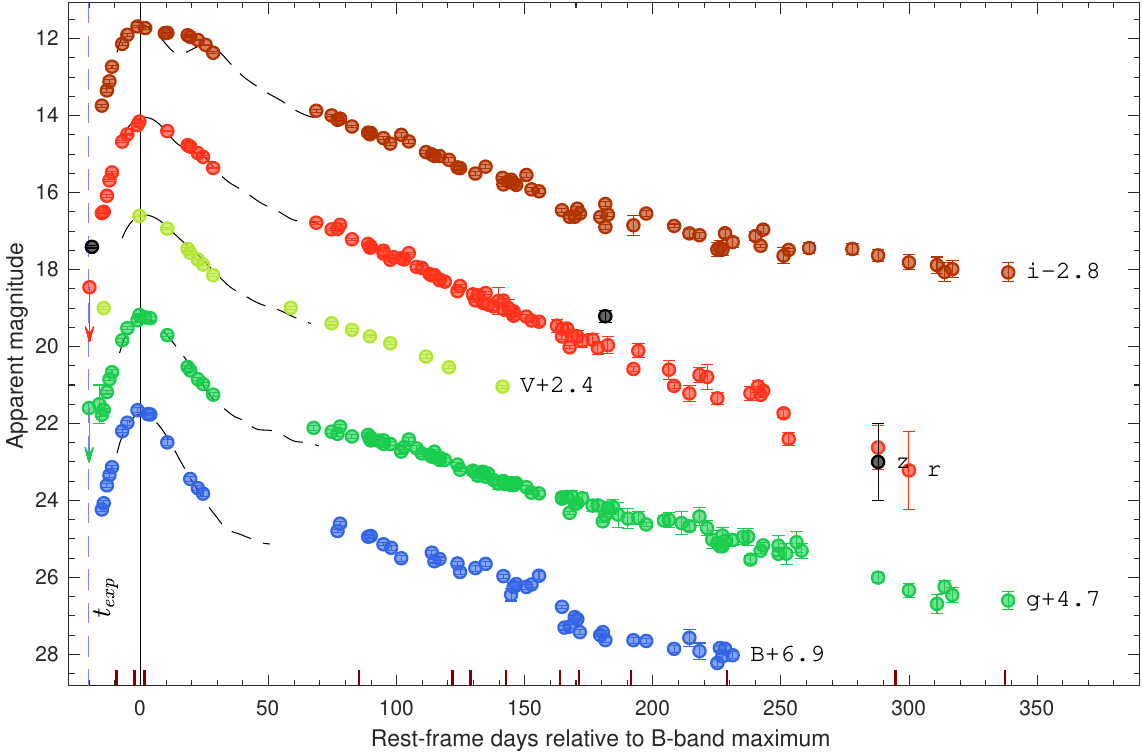}
	\caption{Optical $BgVriz$-band light curves of \sn\ along with ``\texttt{max\_model}'' template light curve fits (black dashed lines) computed by \texttt{SNooPy} \citep{2011AJ....141...19B}. The indicated offsets have been added to the light curves for clarity. Epochs of spectral observations are marked by vertical bars at the bottom, the epoch of $B$-band maximum is indicated with a black vertical line, and the explosion epoch is marked by a blue vertical dashed line. }
\label{fig:all_lc}
\end{figure*}

\begin{table}
\centering
\caption{Key parameters of \sn.}
\label{tab:keyparams}
\begin{tabular}{l|l}
\hline
\hline
Parameters &Values      \\
\hline              
Redshift: $z$ &  $0.00368\pm0.00001$   \\
Distance: $D$ & $19.41\pm3.84$\,Mpc    \\
Distance Modulus: $\mu$ & $31.44\pm0.43$\,mag    \\
Reddening: $\ebv_{tot}$ &  $0.10\pm0.06$\,mag   \\
$B$-band peak magnitude: $ M_{B,max}$ & $-17.09\pm0.50 $\,mag   \\[1.2ex]
\multirow{2}{*}{Explosion time: $t_{exp} \approx t_{first}$}  &  UT 2020-08-03.8  \\
  &  $\rm JD~2, 459, 065.3^{+1.4}_{-3.0}$\,days \\[1.2ex]
\multirow{2}{*}{$B$-band peak time: $t_{B,max}$}  &  UT 2020-08-24.3  \\
  &  $\rm \PeakEpoch\pm0.5$\,days  \\[1.2ex]
$B$-band rise time  &  $20.4^{+3.0}_{-1.5}$\,days \\[1.05ex]
$B$-band decline rate: \dmb  &  $1.35\pm0.09 $\,mag   \\
Color-stretch: \sbv  &  $\gtrapprox0.82$  \\
Synthesized $^{56}$Ni mass: $M_{Ni}$  &  ($0.088\pm0.008$)\,\msun   \\
Host metallicity: $12+\rm{log}_{10}\it{(O/H)}$ &  $8.263\pm0.003$\,dex \\
\hline 
\end{tabular}
\end{table}

\section{Discovery  and distance to \sn} \label{sec:discovery}

ASAS-SN discovered \sn\  on UT 2020-08-08.13 using the ``Cecilia Payne-Gaposchkin'' telescope in South Africa \citep{2016ATel.8566....1B,2017MNRAS.471.4966H} with an apparent $g$-band magnitude of $m_g = 16.79\pm0.08$. Upon reporting to the transient network server (TNS), it received an international astronomical union (IAU) designation SN~2020qxp. Below in Sect.~\ref{sec:tfirst}, we combine the ASAS-SN discovery information with recovered $z$-band photometry obtained with Pan-STARRS (PS) a day prior, as well as last non-detection limits inferred from  ASAS-SN and ATLAS images to obtain a robust constraint of \sn's time of first light (hereafter $t_{first}$).

\begin{figure*}[!ht]
\centering	
\includegraphics[width=0.95\linewidth]{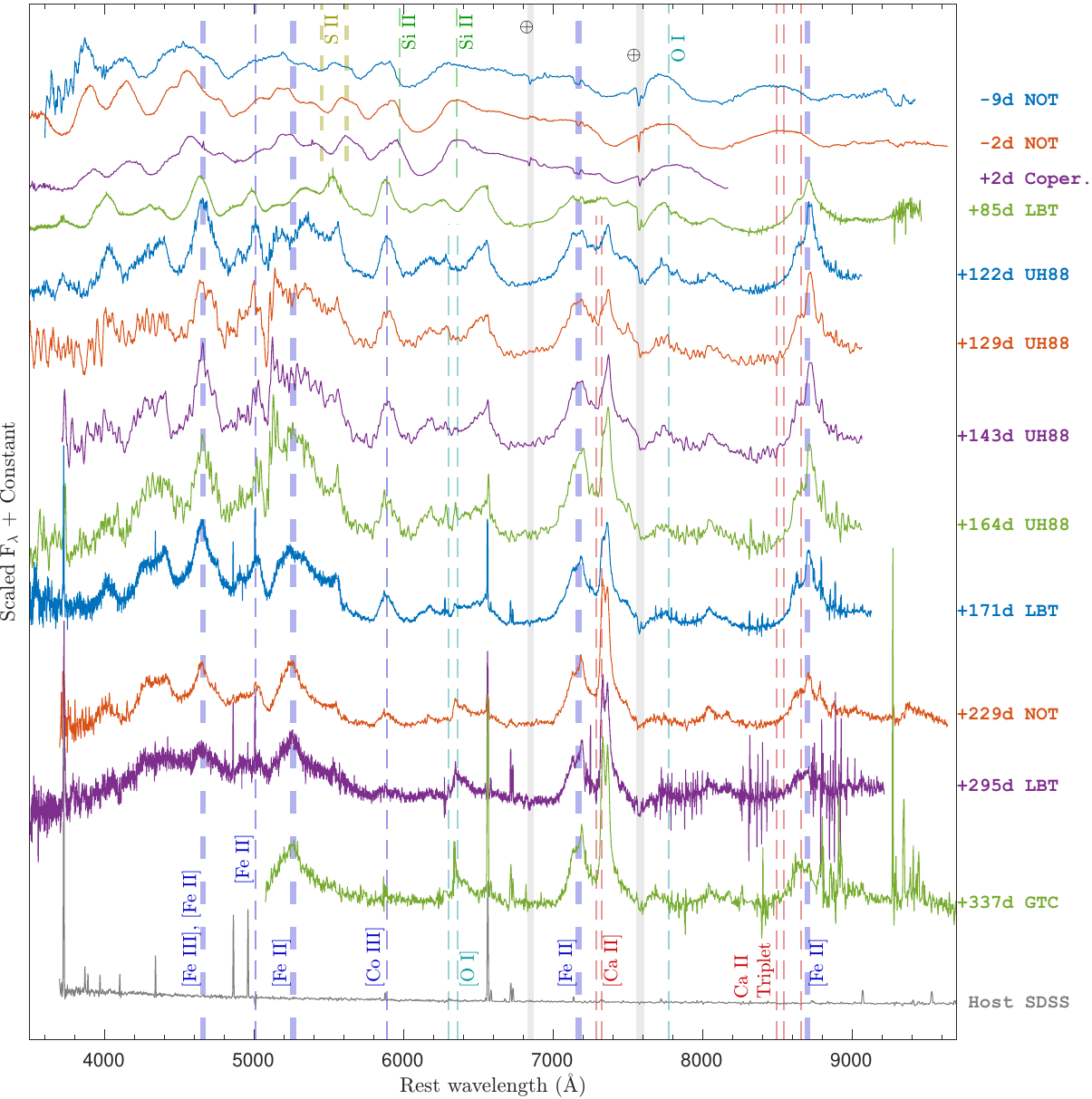}
\caption{Optical spectroscopic time-series of \sn. Positions for telluric oxygen absorption are marked by $\oplus$ symbol and light-grey bands. The vertical dashed lines are at rest-wavelengths for corresponding lines as labeled. Thick dashed lines represent Fe multiplets. An archival SDSS host galaxy spectrum, centered 3\farcs0 away from the SN location, is also included as a reference to show emission lines contamination from the background light in some of the late SN spectra.}
\label{fig:all_spec}
\end{figure*}

With J2000 coordinates of $\alpha=13^{\rm h}10^{\rm m}37\fs79$, $\delta = +36\degr 37\arcmin 43\farcs64$, \sn\ was located in the SBm type %
galaxy \host\  at a heliocentric redshift $z=0.00368\pm0.00001$ \citep{2017ApJS..233...25A}.
The most recent Tully-Fisher measurement to this galaxy implies a distance of  $ 19.41\pm3.84 $ Mpc \citep[Cosmicflows-4;][]{2020ApJ...902..145K}, which we adopt to set the absolute luminosity scale. The Tully-Fisher distance is consistent with the estimate from Cosmicflows-3 Distance–Velocity calculator \citep{2020AJ....159...67K} adopting a smoothed velocity field model \citep{2017ApJ...850..207S}. 
Key parameters of \sn\ are listed in Table~\ref{tab:keyparams}.

\section{Observations and data reductions} \label{sec:obsv}

\begin{table}
\centering
\caption{Journal of spectroscopic observations.}
\label{tab:speclog}
\begin{tabular}{llll}
\hline
\hline
Date (UT)  &JD$-$ &Phase\tablefootmark{a} & Telescope \\
           &2,459,000 & (days) & Instrument\\
\hline              
2020-08-14.90 &  076.40 & $-$9  & NOT/ALFOSC       \\
2020-08-21.87 &  083.37 & $-$2  & NOT/ALFOSC       \\
2020-08-25.84 &  087.34 &   +2  & Copernico/AFOSC \\
2020-11-18.00 &  171.50 &  +85  & LBT/MODS         \\
2020-12-18.24 &  201.74 & +116 & TNG/LRS          \\
2020-12-24.61 &  208.11 & +122  & UH88/SNIFS      \\
2020-12-31.61 &  215.11 & +129  & UH88/SNIFS      \\
2021-01-14.59 &  229.09 & +143  & UH88/SNIFS      \\
2021-02-04.58 &  250.08 & +164  & UH88/SNIFS       \\
2021-02-12.00 &  257.50 & +171  & LBT/MODS         \\
2021-04-11.01 &  315.51 & +229  & NOT/ALFOSC       \\
2021-06-16.00 &  381.50 & +295  & LBT/MODS         \\
2021-07-28.91 &  424.41 & +337  & GTC/OSIRIS       \\
\hline
\end{tabular}
\tablefoot{
\tablefoottext{a}{Rest-frame days relative to the epoch of $B$-band maximum, i.e., \PeakEpoch.}
}
\end{table}

\begin{figure}[!ht]
\centering
\includegraphics[width=\linewidth]{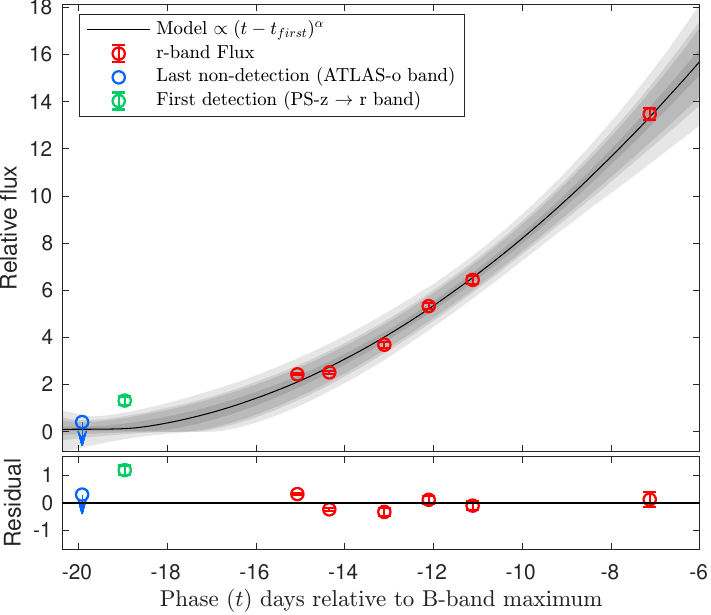}
\hspace*{-0.6cm}
\includegraphics[width=1.15\linewidth]{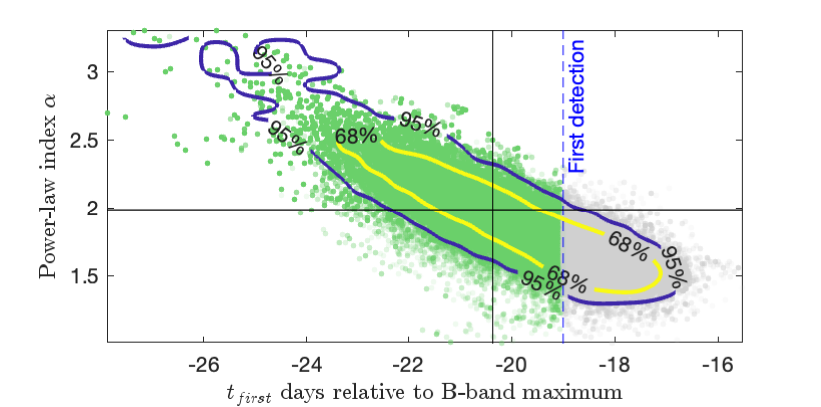}
\caption{Top panel: The early-time $r$-band light curve of \sn. This also includes the first photometric detection of \sn\ made by PS in the $z$~band, which has been S-corrected to the SDSS $r$-band system. Also plotted is the ATLAS $o$-band previous non-detection limit. Over-plotted the $r$-band light curve is the best MCMC-fit of a single power-law function (solid curve) with inferred probability limits (shaded region). 
 Bottom panel: The 2-D probability density of MCMC sample between the fit parameters -- time of first light $t_{first}$ and power-law index $\alpha$. Having a confirmed detection at $-19$~days (vertical blue dashed line), only the region in `green' is valid parameter space. The contours correspond to  68\% and 95\% confidence intervals. The solid black lines indicate the mean of the MCMC sample for parameters $t_{first}=-20.4$\,days (vertical) and $\alpha=1.99$ (horizontal).
 }
\label{fig:tfirst}
\end{figure}

We initiated a multi-band photometric and spectroscopic campaign shortly after the initial discovery. Our follow-up spanned over a year and in addition to combined resources of ASAS-SN, and the 2.6\,m Nordic Optical Telescope (NOT) on La Palma, Spain, through the NOT Un-biased Transient Survey (NUTS2; \citealt{Holmbo2019})\footnote{\href{https://nuts.sn.ie/}{https://nuts.sn.ie/}}, optical photometric data were obtained with the  2.0\,m Las Cumbres Observatory Global Telescope network \citep[LCOGT;][]{2013PASP..125.1031B}, the 0.6\, and 0.8\,m telescopes at Post Observatory SRO (CA, USA) and Post Observatory Mayhill (NM, USA), and the 0.9\,m Schmidt telescope at Asiago, Italy. Our photometry is also complemented by photometry from the ZTF survey. All of the broadband imaging data were reduced following standard procedures.

\begin{figure}[!ht]
\centering
\includegraphics[width=0.95\linewidth]{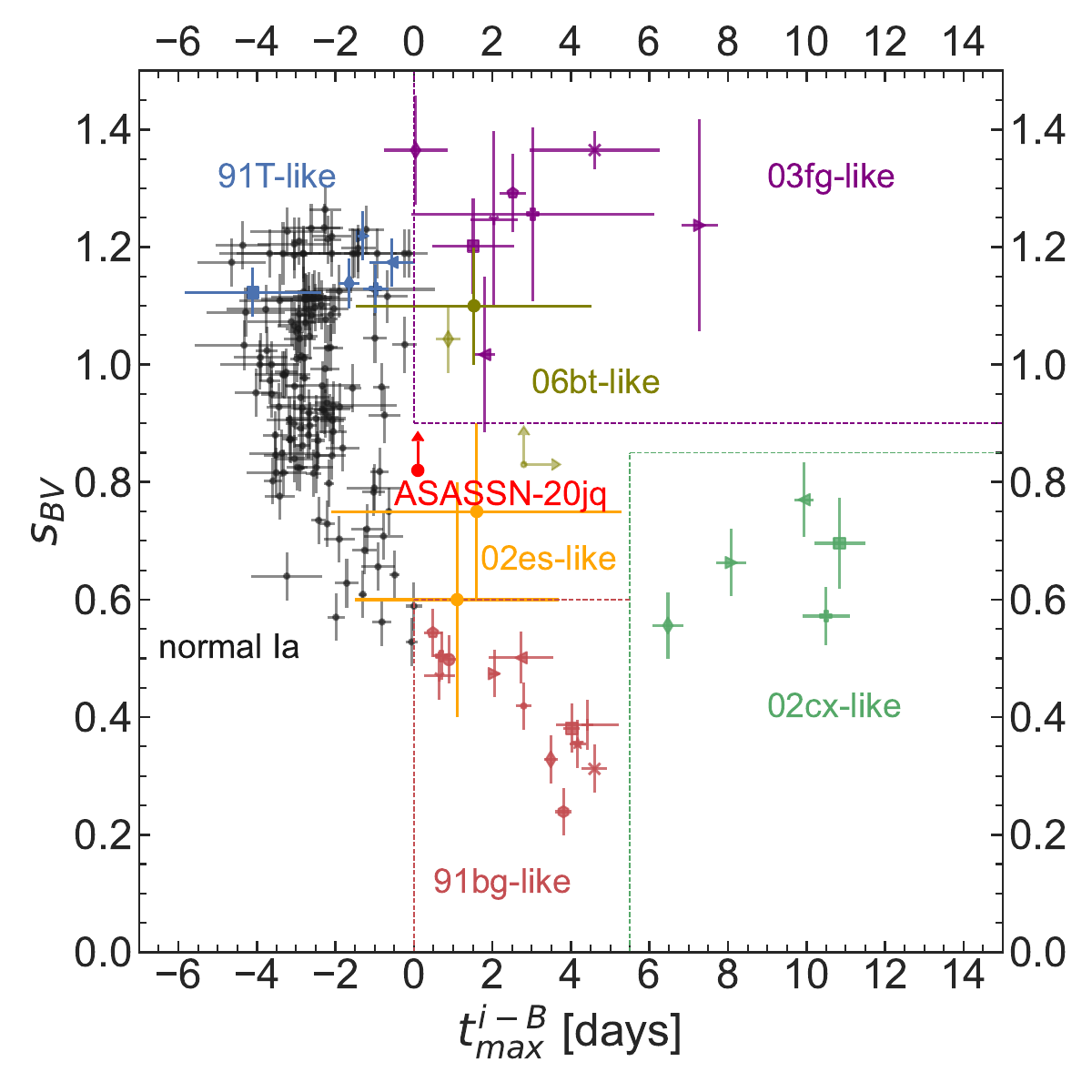}
\caption{Sub-typing of SNe~Ia based on light curve properties as per \citet{Ashall2021}. Specifically, the color-stretch \sbv\ is plotted vs. $t^{i-B}_{max}$, defined as the difference between the epoch of $i$- and $B$-band light curve maximum.}
\label{fig:t_i_B}
\end{figure}

Prior to computing photometry, host-galaxy subtraction was performed on all science images. To do so, a common set of $BgVriz$-band images containing the field of the host galaxy were obtained two years after discovery by the Aarhus-Barcelona cosmic FLOWS project\footnote{\href{https://flows.phys.au.dk/}{https://flows.phys.au.dk/}} using the  LCOGT 2\,m telescope. The template images were subtracted from the science images using the \texttt{SUBPHOT} package which is an extension of the differential photometry pipeline \texttt{DIFFPHOT}.\footnote{\href{https://pypi.org/project/diffphot/}{https://pypi.org/project/diffphot/}} The \texttt{SUBPHOT} pipeline first analyzes image quality, optimizes the images as a pre-requisite for template subtraction, and aligns the science images with templates, following which it employs the \texttt{HOTPANTS} software \citep{2015ascl.soft04004B} to perform the optimal image subtraction. 

 Differential PSF photometry of \sn\ was computed using \texttt{DIFFPHOT} pipeline on the sequence of template subtracted science images. Photometry was calibrated using
 common field stars from PS photometric catalog \citep{2020ApJS..251....7F}.
 Prior to computing zeropoints for calibration, the PS catalog photometry is converted to the Johnson $BV$ and Sloan $griz$ photometric system using the color relations of  \citet{2012ApJ...750...99T}. 
 The $BgVriz$-band light-curves of \sn\  extending between $-$15~days to $+$339~days relative to the epoch of $B$-band maximum are plotted in Fig.~\ref{fig:all_lc}. In this paper, unless otherwise stated, the time of $B$-band maximum brightness ($t_{B,max}$ in Table.~\ref{tab:keyparams} and Sect. \ref{sec:earlyphot}) is used as the reference epoch. The photometry is listed in Table.~\ref{tab:photsn}, along with ZTF photometry, which leads to more densely sampled $g$- and $r$-band light curves.  

Thirteen low-resolution optical spectra of \sn\  were obtained using multiple facilities including the  NOT (+ ALFOSC), the 8.4\,m Large Binocular Telescope  \citep[LBT, + MODS;][]{2010SPIE.7735E..0AP}, the 2.2\,m Hawai'i  telescope  \citep[+ SNIFS;][through SCAT survey \citealp{2022PASP..134l4502T}]{2002SPIE.4836...61A}, the  1.8\,m Copernico telescope (+ AFOSC), the 3.6\,m Telescopio Nazionale Galileo (TNG, +LRS), and the 10.4\,m  Grand Telescopio Canarias (GTC, + OSIRIS). The journal of spectroscopic observations is provided in Table~\ref{tab:speclog}. The spectroscopic data were reduced and calibrated following standard procedures. %

The spectroscopic evolution of \sn\ is shown in Fig.~\ref{fig:all_spec} with the key features labeled.\footnote{The spectra are electronically available at the Weizmann Interactive Supernova Data Repository (\href{https://www.wiserep.org//object/}{WISeREP}; \citealt{Ofer2012}).} The first 3 spectra obtained on $-$9~days, $-2$~days, and $+$2~days cover the so-called photospheric phase of the SN evolution,  while the remaining spectra extending between $+$85~days and $+$337~days provide a nebular phase window into the inner ejecta of the progenitor star \citep[see][and below]{Hoeflich2021}.

\section{Broadband photometry} 
\label{sec:photometry}

\begin{table*}
\centering
\caption{Peak time, apparent, and absolute magnitudes for each filter.}
\label{tab:peakvalues}
\begin{tabular}{cccccc}
\hline
\hline 
Filter     & 
$t_{max}$  & 
$t_{max}-t_{B,max}$ (days)& 
$m_{max}$ (mag)& 
$m_{max}$ (mag)& 
$M_{max}{~}^b$ (mag)
\\& 
($\rm JD-2459000$) & 
Spline &
Spline&
SNooPy& Spline\\		
\hline		
$B$  & $85.8 \pm 0.8$       &$\cdots$         &  $14.76\pm0.06  $ &$14.65\pm0.03$& $-17.09\pm0.50$ \\
$V$  & $~88.4 \pm 3.0^a$  &$~2.6\pm3.1^a$   &  $14.09\pm 0.08 $ &$14.15\pm0.03$& $-17.66\pm0.48$ \\
$g$  & $86.6 \pm 0.9$       &   $~0.8\pm1.2~$ &  $14.50\pm 0.03$ &$14.35\pm0.02$& $-17.31\pm0.49$\\
$r$  & $~87.3 \pm 2.0^a$    &$~1.5\pm2.2^a$   &  $14.06\pm 0.23 $ &$14.00\pm0.03$& $-17.65\pm0.52$\\
$i$  & $85.7 \pm 1.0$       &$-0.1\pm1.2~$    &  $14.49\pm 0.03$ &$14.23\pm0.03$& $-17.15\pm0.45$ \\
\hline
\end{tabular}
\tablefoot{
\tablefoottext{a}{The larger uncertainties for $V$ and $r$ band measurements are due to poor sampling of the light curve near the peak.}\\
\tablefoottext{b}{Quoted uncertainties account for errors in the estimated apparent peak magnitude, distance, and extinction.}
}
\end{table*}

\subsection{Photospheric phase optical light curves}
\label{sec:earlyphot}

The multi-band photometry in Fig.~\ref{fig:all_lc}) shows the bell-shaped light curves ubiquitous to thermonuclear SNe. Our coverage nicely samples in most bands the rise to maximum and the post-maximum declining phases. At some time between  $+$30~days to  $+$60~days when \sn\ was located behind the Sun, the transition occurred to the linear declining phase powered by $^{56}$Co to $^{56}$Fe radioactive decay energy deposition, which is followed for several hundred days.

Generic SNe~Ia and overluminous 1991T-like objects exhibit double-peaked $i$-band light curves.  \sn\ does not show a distinct secondary maximum, but rather, its $i$-band light curve exhibits a flatter profile over $\sim 20$ days near the maximum. This morphology is typical of many ``peculiar'' subtypes including 1991bg-like \citep{Krisciunas2009}, 2002cx-like \citep{Foley2013}, 2003fg-like \citep{2020ApJ...895L...3A}, and 2002es-like \citep[e.g.,][see their Fig.~4]{Burke2021} SNe~Ia.

In Fig.~\ref{fig:all_lc} we also over-plot along with each observed  light curve their corresponding  best-fit template light curve computed using  the \snoopy\ 
 [SuperNovae in object-oriented Python; \citealt{2011AJ....141...19B}] code. As \sn\ is a peculiar supernova, we make use of the so-called ``max\_model'' function to compute these template fits \citep[see][Eq.~5]{Stritzinger2010}. Inspection of the template fits indicates reasonable agreement in all but the $i$-band, which is due to \sn\ lacking of a distinct secondary maximum.

In addition to template light curve fits, the early light curves were fit with Gaussian Process (GP) spline function to obtain estimates of the peak magnitude and the epoch of maximum. These results along with those obtained from template fitting are summarized in Table~\ref{tab:peakvalues}.
The spline fits provide $m_{B,max}=14.76\pm0.03 $\,mag, a time of peak $ t_{B,max}=\rm\PeakEpoch\pm0.5 $ days, and a light-curve decline-rate parameter of $\dmb = 1.35\pm0.09 $\,mag. The corresponding peak absolute magnitude (corrected for reddening, see below) is $ M_{B,max}=-17.09\pm0.50 $\,mag, which is $\sim$2~mag fainter than normal SNe~Ia with a similar decline rate.

\subsection{Estimating the epoch of first light and the rise time}
\label{sec:tfirst}

The first detection of \sn\ was made from a PS survey $z$-band image taken on UT~2020-08-05.25, which corresponds to $-$19 days prior to the epoch of $B$-band maximum. Previous, rather shallow non-detection limiting $g$-band and $o$-band apparent magnitudes of 16.9 and 18.46 are estimated from ASAS-SN and ATLAS survey images taken on  UT~2020-08-04.14 and UT~2020-08-04.29, respectively (see Table~\ref{tab:photsn}). Comparing the non-detection and discovery epoch to the epoch of $B$-band maximum indicates a minimum rise-time for \sn\ of $\sim 19.4\pm0.7$\,days. This value is typical for normal SNe~Ia \citep[e.g.,][] {2017hsn..book..317T}.

\begin{figure}[!t]
\centering
\hspace*{-4.5mm}
\includegraphics[width=1.12\linewidth]{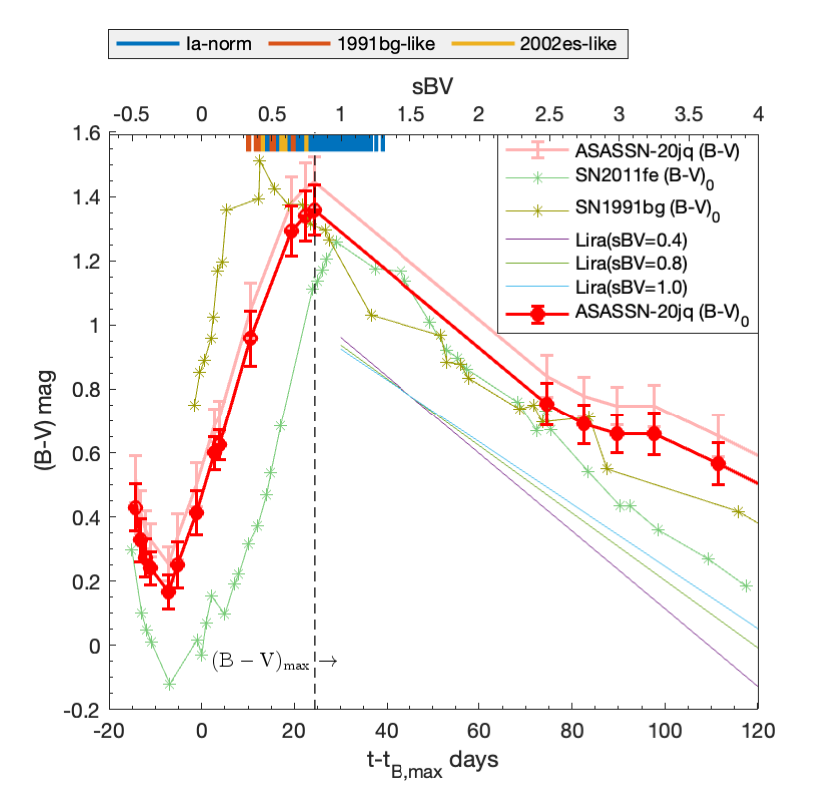}
\caption{Apparent $(B-V)$ and intrinsic $(B-V)_{0}$ color evolution of \sn. The intrinsic color evolution of the normal SN~2011fe \citep{2012JAVSO..40..872R},  the low-luminosity SN~1991bg \citep{1996MNRAS.283....1T}, and the Lira relation for different $s_{BV}$ values \citep{Burns2014} are also plotted. To construct the color-curve of \sn, $B$- and $V$-band light curves were interpolated with low-order splines which were then evaluated at the same epochs. 
The vertical dashed line marks the lower limit of the time of \bv\ maximum of \sn\ as the data only captures the rising part of the color curve. The vertical bars at the top of the figure indicate the range of \sbv\ values for the SNe~Ia-normal, 1991bg-like, and 2002es-like samples populating the luminosity vs. \sbv\ diagram presented in Fig.~\ref{fig:dm15sbv}.}
\label{fig:bv_color}
\end{figure}

\begin{figure}[!t]
\centering
\includegraphics[width=0.97\linewidth]{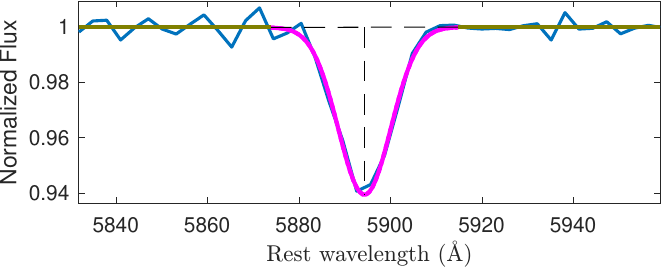}
\caption{\Nai~D lines at the rest wavelength of the host galaxy. The $-2.4$ days spectrum is normalized by the local continuum. The D1 and D2 lines are unresolved in this low-resolution spectrum.
The estimated $EW$ for the \Nai~D profile is $0.69\pm0.04$\,\AA\ and the corresponding $E(B-V)_{host}$ is $0.091\pm0.020$\,mag.}
\label{fig:NaID}
\end{figure}

To more accurately estimate the rise time, we fit the rising light curve with a power law function to estimate $t_{first}$, the time of first light (Fig.~\ref{fig:tfirst}). 
As a first step, the  PS $z$-band natural system recovered detection photometry was S-corrected \citep{Stritzinger2002}  to the SDSS $r$~band  so that it can be fit with our later $r$-band light curve. This was accomplished by computing 
synthetic photometry convolving the respective broadband filter response functions with an appropriate spectral energy distribution  (SED) of \sn.  
Due to the lack of observed spectra of \sn\ at $-19$~days, we assume a range of possible SEDs and assign the results as limits of the uncertainty associated with the transformed photometric fluxes. 
To estimate the upper limit, we used the earliest publicly available SN~Ia spectrum, that is SN~2017cbv  at $-18.3$~days \citep{2017ApJ...845L..11H}, while to estimate a lower limit we use the $-9$~days spectrum of \sn.
{A black-body SED was also considered, adopting a temperature of $\sim1.0\times10^4$,K at approximately +1.5 days after $t_{first}$ of SN~2011fe \citep{2018ApJ...858..104Z}. However, this value falls within the range assumed based on the spectra of \sn\ and SN~2017cbv.}

\begin{figure*}[!ht]
\centering 
\includegraphics[width=0.49\linewidth]{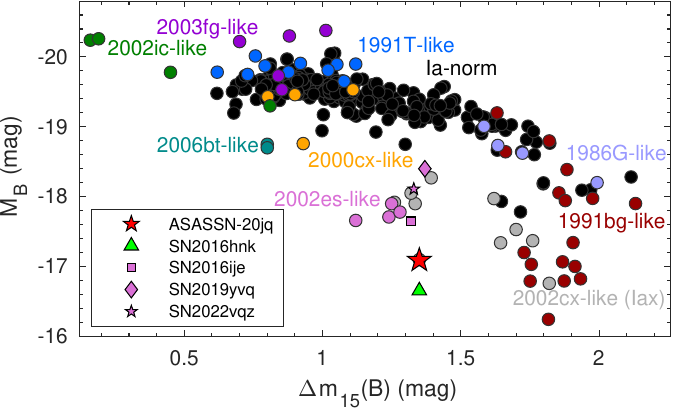} 
\includegraphics[width=0.49\linewidth]{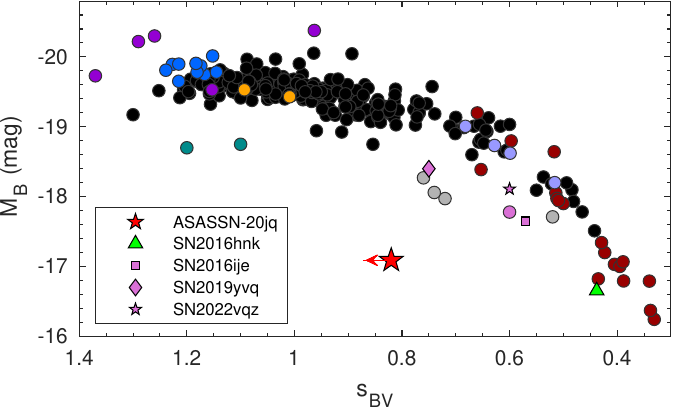}
\caption{Absolute magnitude versus light curve shape diagrams populated with the Carnegie Supernova Project SN~Ia sample \citep{Contreras2010,Stritzinger2011,Krisciunas2017,Ashall2021,2022ApJ...938...47P}, along with members of known SN-Ia subtypes, a handful of  2002es-like  objects, and \sn. References for the expanded comparison sample are provided within Sect.~\ref{sec:dm15outlier}.
 The left panel relates the peak absolute $B$-band magnitude ($M_B$) to the light curve shape parameter  \dmb\ \citep{Phillips1993}, while $M_B$ is plotted in the right panel vs. the color-stretch parameter \sbv\ \citep{Burns2014}.} 
\label{fig:dm15sbv}
\end{figure*}

With the transformed $z$- to $r$-band PS photometry, a power-law function was fitted to the early rising $r$-band light curve of \sn\ using the Markov chain Monte Carlo (MCMC) package \texttt{MCMCSTAT} \citep{haario2006dram}, which employs adaptive Metropolis samplers \citep{haario2001adaptive}.
The MCMC fit was performed using all early $r$-band data points extending up to $-$7~days \citep{Conley2006,Ganeshalingam2011}. A preliminary least-squares fit was first conducted to obtain an initial power-law model, which was then used as the starting parameter for the MCMC fitting process.

The top panel of Fig.~\ref{fig:tfirst} shows the early rising $r$-band light curve of \sn\ along with the last ATLAS $o$-band non-detection limit and the best-fit MCMC model. The bottom panel presents the 2-D probability density of the MCMC sample for the power-law index parameter $\alpha$ and $t_{first}$.
The best-fit corresponds to an $\alpha=1.99^{+0.59}_{-0.51}$ and $t_{first}$ of $\rm JD~2, 459, 065.3^{+1.4}_{-3.0}$ (UT 2020-08-03.8), which is 1.5~days before the epoch (UT 2020-08-05.3) of the recovered PS $z$-band detection and 
  $-20.4^{+1.5}_{-3.0}$\,days relative to $t_{B,max}$.
The inferred value of $\alpha$ is fully consistent with the expanding ``fireball'' model characterized by $\alpha = 2$.

However, close inspection of the PS $z$-band discovery photometry transformed to the SDSS $r$~band obtained on $-19$~days indicates that it is a 3-$\sigma$ outlier relative to the power-law fit.  This suggests an excess of flux by $2.3\pm1.3$\,mag relative to the expanding fireball model, at least during the first 1.4 days of the estimated $t_{first}$. 
{It is worth emphasizing that, although the early bump is statistically significant, there is inherent uncertainty in the S-corrected photometric data point due to the assumed SED at $-19$ days. The range of SEDs considered includes those of SNe~2011fe and 2017cbv, which are not representative of 2002es-like SNe and may differ from that of \sn\ at such an early phase. Additionally, SN~2017cbv exhibited an early flux excess, which could further influence the photometric transformation.}

\subsection{Late phase light curve}

The high-cadence photometry of \sn\ extending beyond $+$50 days enables for a robust measurement of the post maximum $^{56}$Co $\rightarrow$ $^{56}$Fe powered light curve decline rate.  Normal SNe~Ia typically exhibit $B$- and $V$- linear decline rates of $\approx 1.4$ mag per 100 days  \citep[e.g.,][]{Lair2006,Stritzinger2007,Leloudas2009}.

A best-fit linear function to the  $B$-band light curve of \sn\ between +75~days to $+$230 days reveals a decline rate of $2.3\pm0.1$ mag per hundred days, while similar fits to the  $r$- and $i$-band data give decline rates of $2.8\pm0.1$  and $2.2\pm0.1$ mags per hundred days respectively. We also note that the post +250 days $i$-band light curve exhibits a slower decline relative to the continued linear decline exhibited by the $g$ and $r$ bands (see Fig.~\ref{fig:all_lc}), indicating a shift in the percentage of total flux towards infrared bands. Similar late-time flattening in $I$-band was also seen in SN~2011fe \citep{2013NewA...20...30M}.

\subsection{Photometric classification}

For SNe~Ia sub-types exhibiting double-peaked light curves (i.e., normal and 1991T-like SNe~Ia), the $i$-band light curve reaches peak luminosity typically 3-5 days prior to  epoch of $B$-band maximum. On the other hand, for subtypes without prominent double-peaked light curves (i.e., 1991bg-like, SNe~Iax, and 2003fg-like), the epoch of $i$-band maximum occurs after that of the $B$ band \citep{Krisciunas2009,Stritzinger2015,Ashall2021}.  
\citet{Ashall2021} recently leveraged these empirical findings to provide a photometric-based means to subtype SNe~Ia. By comparing the difference between the epochs of peak  $i$- and $B$-bands ($t^{i-B}_{max}$), and the light curve color-stretch parameter ($s_{BV}$)\footnote{The color-stretch parameter, \sbv, is the number of days for the \bv\ color to reach its maximum value normalized by 30 days \citep{Burns2014}. This epoch marks a transition in the ionization state of the photosphere as  \ion{Fe}{iii} recombines to \ion{Fe}{ii} \citep{Burns2014,Wygoda2019}.} they find clear groupings of known SNe~Ia subtypes. We recreate the results of \citet{Ashall2021}  in Fig.~\ref{fig:t_i_B}.
Our spline fits to the photometry of \sn\ indicate $t^{i-B}_{max} = 0.1$ days. This places \sn\ within a sparsely populated region of the figure, with no overlap between the major SN~Ia subclasses.

\subsection{Colors and reddening estimation}
\label{sec:reddening}

Figure~\ref{fig:bv_color} shows the apparent \bv\ color evolution of \sn\ extending from the time observations were commenced through $+$100\,d. Also plotted for comparison  are the intrinsic $(B-V)_0$ color-curves associated with the normal type~Ia SN~2011fe and the subluminous SN~1991bg. Inspection of the color evolution of  \sn\  reveals a dearth of coverage, and as a result, the peak of the \bv\ color is not captured. We therefore obtain a lower limit on  the color-stretch parameter of \sbv $> 0.82$. This value is consistent with  $\sbv = 0.76\pm0.08$,  as determined by the \snoopy\ template light curve fits shown in Fig.~\ref{fig:all_lc}. 

Inspection of Fig.~\ref{fig:bv_color}  reveals the apparent \bv\ colors of \sn\ are redder than those of SN~2011fe by $\approx 0.1$ mag at maximum, and typically between  $\sim 0.1-0.3$~mag over the entire photospheric phase. The peak color difference between \sn\ and SN~2011fe is fully consistent with those of other 2002es-like SNe~Ia  \citep[see, e.g.,][]{Burke2021}.  
As  the coverage of \sn\  between $+$30 days to $+$90~days is nearly nonexistent, we are unable to  estimate the host reddening via the Lira relation, though the limited observations in hand do imply  \sn\ suffers minimal host reddening.  

 We next turn to the \Nai~D absorption lines detected in spectra obtained at $-2$\,days and $+85$\,days to ascertain the host-reddening of \sn. 
Using the higher signal-to-noise ratio $-2$\,days spectrum (see Fig.~\ref{fig:NaID}), the pseudo-equivalent width ($pEW$) of the \Nai~D feature is measured to be $pEW(\Nai~D)=0.69\pm0.04$\,\AA. 
Adopting the empirical-based relation prescribed by \citet{Poznanski2012},
this translates to $\ebv_{host}=0.091\pm0.020$\,mag. 
Similarly, we measure the \Nai~D profile in $+85$\,days spectrum and obtain a $\ebv_{host}$ of $0.064\pm0.037$\,mag. However, we discard the later measurement due to its low signal-to-noise (S/N) spectrum.
\cite{2013ApJ...779...38P} further studied the relation between $pEW(\Nai~D)$ and reddening using an expanded data sample of high-resolution spectroscopy of SNe~Ia, as well as from measurements in the  Milky Way. 
They found a dispersion in the Milky Way correlation of $ \sim68\% $ instead of $ \sim20\% $ as suggested by \cite{Poznanski2012}. Therefore, with the addition of an $\sim68\% $ uncertainty to the reddening estimated above we arrive at our final adopted reddening value of $ \ebv_{host} = 0.091\pm0.062 $\,mag. 
This value is similar to the value inferred from the  \Nai~D vs. reddening relation calibrated based on the broad-band colors of stripped-envelope SNe \citep{Stritzinger2018,2023ApJ...955...71R}, which implies a somewhat higher $E(B-V)_{host} \approx 0.17$ mag. 

Combining our adopted host reddening with the
 Milky Way reddening measured along the line-of-sight  to \sn\ corresponding to $\ebv_{MW}=0.0096$ mag \citep{2011ApJ...737..103S}, yields a total $E(B-V)_{tot} = 0.10\pm0.06$\,mag. Upon adopting the standard total-to-selective absorption value of $R_V=3.1$, it provides a $V$-band extinction value of  $A^{tot}_{V} = 0.31\pm0.19$ mag. 

\subsection{\sn\ is a significant outlier in luminosity-width relation}
\label{sec:dm15outlier}

The left panel of Fig.~\ref{fig:dm15sbv}  shows the SNe~Ia luminosity-width relation (or the \citet{Phillips1993} relation) populated with normal SNe~Ia from the Carnegie Supernova Project \citep{Hamuy2003,2019PASP..131a4001P}, along with members of known SN~Ia subtypes. The bright end of the distribution is populated by bright normals,  overluminous 1991T-like \citep{Filippenko1992b,Phillips1992,2022ApJ...938...47P}, 2002ic-like \citep{Hamuy2003},  and 2003fg-like \citep{Howell2006,Ashall2021} subtypes. The faint end of the distribution is populated by the so-called  transitional  SNe~Ia \citep[e.g., iPTF~13ebh;][]{Hsiao2015}, the low-luminosity  1986G-like \citep{Phillips1987} subtype, and the subluminous 1991bg-like subtype. Additional peculiar subtypes including 2000cx-like \citep{Li2001}, 2002cx-like \cite[Iax,][]{Li2003}, and  2002es-like objects are also included.  Specific 2002es-like objects \citep{Ganeshalingam2012} are PTF10ops \citep{Maguire2011}, iPTF14atg \citep{Cao2015}, and SN~2016ije  \citep{Li2023}. These are further complemented by other objects with commonalities to 2002es-like SNe including SN~2006bt \citep{Foley2010,Stritzinger2011}, SN~2006ot \citep{Stritzinger2011}, SN~2019yvq  \citep{Burke2021}, SN~2022vqz  \citep{Xi2024}, and SN~2016hnk \citep{Galbany2019,Jacobson2020}.

In the left panel, Fig.~\ref{fig:dm15sbv} \sn\ is located  $\approx2.5$\,mag below the Phillips relation, sandwiched between a handful of 2002es-like and 2002cx-like SNe being more luminous, and SN~2016hnk being less.
The $\Delta m_{15}(B)$ parameterization is known to be degenerate for low-luminosity, fast-declining SNe~Ia (e.g., 1991bg-like SNe).
Turning to the right panel, which instead relies on the more robust color-stretch parameter \sbv\ \citep{Burns2014}, we see that \sbv\ parameter does a good job in significantly reducing the scatter along the Phillips relation, particularly for the fast declining SNe~Ia.
Among other SNe, the reduced dispersion also brings SN~2016hnk close to the distribution of 1991bg-like SNe~Ia. 
This is also the case of SN~2002es and other similar 2002es-like objects such as SNe~2016ije \citep{Li2023}, 2019yvq \citep{Miller2020}, and 2022vqz \citep{Xi2024}.
Remarkably, \sn\ appears immune to the \sbv\ standardization as well, since it remains 2.5\,mag below the luminosity-color-stretch relation, and it is the only significant outlier in the comparison sample.

\begin{figure}[!t]
\centering
\includegraphics[width=0.90\linewidth]{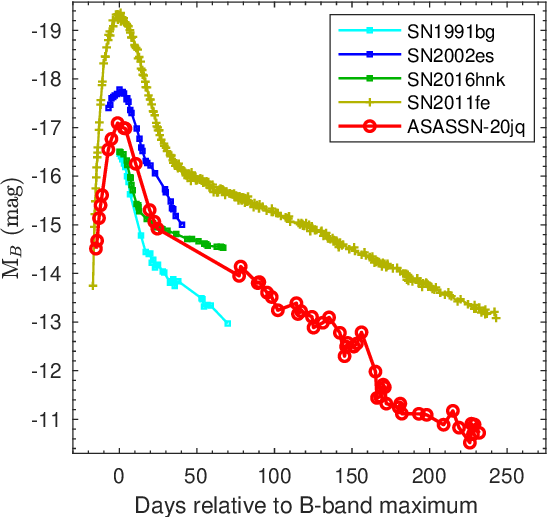}
\caption{Absolute $B$-band magnitude light curves of \sn\  compared with the normal SN~2011fe, the subluminous SN~1991bg,  as well as the peculiar SNe~2002es and 2016hnk.   The adopted distance, $A^{tot}_B$ extinction, and references of the comparison sample are as follows: SN~2011fe -- 6.79 Mpc, 0.93 mag \citep{2012JAVSO..40..872R,2013A&A...549A..62P}, SN~1991bg -- 17.0~Mpc, 0.0\,mag \citep{Leibundgut1993}, SN~2016hnk -- 68.35 Mpc, 1.48 mag, \citep{Galbany2019}, SN~2002es -- 73.2~Mpc, 0.00 mag \citep{Ganeshalingam2012}.}
\label{fig:abs_mag}
\end{figure}

\subsection{Absolute magnitude and  bolometric light curves}

Figure~\ref{fig:abs_mag} compares the  absolute $B$-band magnitude light curves of \sn\ with a comparison sample 
consisting of the normal SN~2011fe \citep{2012JAVSO..40..872R}, as well as the low-luminosity SNe~1991bg \citep{1996MNRAS.283....1T},  2002es \citep{Ganeshalingam2012} and 2016hnk \citep{Galbany2019}. The top panel demonstrates the peak magnitude of  \sn\ is sandwiched between SNe~2002es and 2016hnk, and that its post maximum phase follows a standard decline rate. 

\begin{figure}[!t]
\centering	\includegraphics[width=0.97\linewidth]{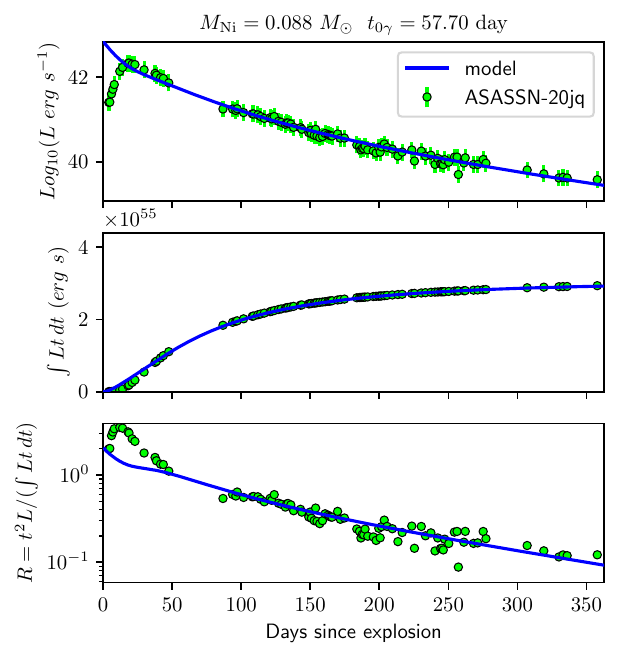}
\caption{Modeling the UVOIR bolometric light curve with a radioactive $^{56}$Ni decay model (solid lines). The UVOIR  light curve of \sn\ (top panel), its time-weighted integrated luminosity (middle panel), and the ratio $t^2L/(\int L t~dt)$ (bottom panel) which is defined independent of the \nickel\ mass are shown. Phase is in days since the adopted time of first light, which is 20.4 days before the epoch of the $B$-band maximum (see Sect.~\ref{sec:tfirst}).}
\label{fig:bol_nimass}
\end{figure}

To estimate the quantity of \nickel\ synthesized during \sn's disruption, we constructed and modeled a UVOIR bolometric light curve. 
First, the broad-band optical light curves were interpolated using Gaussian Process spline functions. A reddening correction was applied adopting $E(B-V)_{tot} = 0.10$~mag. Next, using the bolometric function contained within \texttt{SNooPy}, 1991bg-like spectral templates from \citet{Nugent2002} were color-matched to the reddening-corrected broad-band colors of \sn.\footnote{The Nugent templates  are electronically available at \url{https://c3.lbl.gov/nugent/nugent_templates.html}.} To account for flux beyond the spectral range of the spectral templates we (i) extended the SED bluewards of the atmospheric cutoff by linearly extrapolating to zero flux at 2000\,\AA\ and, (ii) extended the SED redwards of the $i$-band spectral range using a Rayleigh-Jeans tail extending to 20,000\,\AA.  Upon integration of the color-corrected spectral templates, the integrated flux values were converted to luminosity adopting the distance quoted in Table~\ref{tab:keyparams}.
The resultant UVOIR light curve is plotted in the top panel of Fig.~\ref{fig:bol_nimass}.

\begin{figure*}%
\centering  \includegraphics[height=0.85\linewidth]{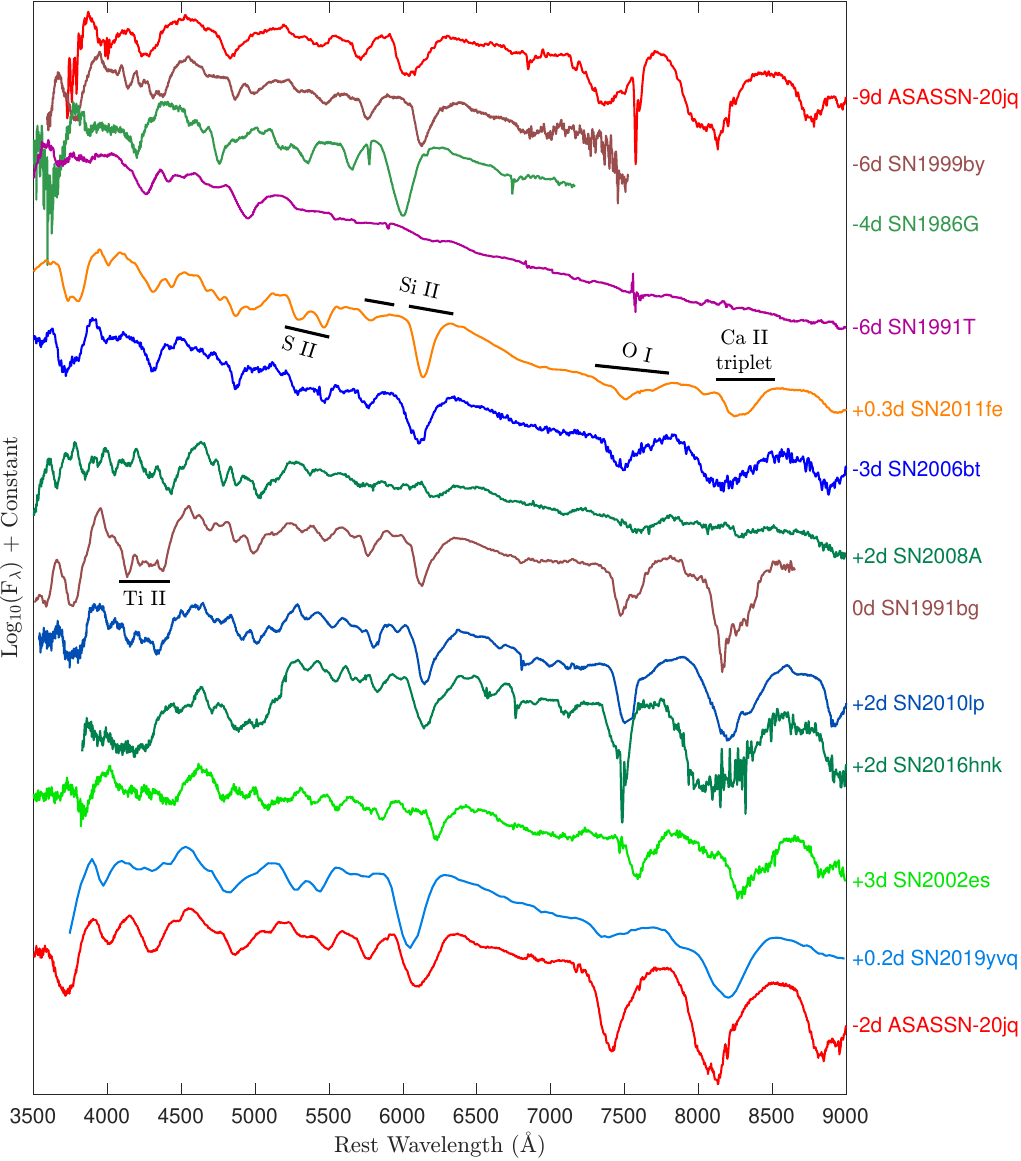}
\caption{Photospheric spectra  of \sn\ compared with similar epoch spectra of other SN~Ia types with key spectral features marked and labeled. 
The comparison sample includes subluminous SNe 1991bg, 1999by, and 2016hnk, the normal SN 2011fe, the overluminous SN 1991T, the Type Iax SN 2008A, and SN 2002es, along with several commonly referenced objects in our overall comparison, namely SNe 2006bt, 2010lp, and 2019yvq. See text for references.}
\label{fig:spec_comp_peak}
\end{figure*}

\begin{figure}[!ht]
	\centering
\includegraphics[width=0.98\linewidth]{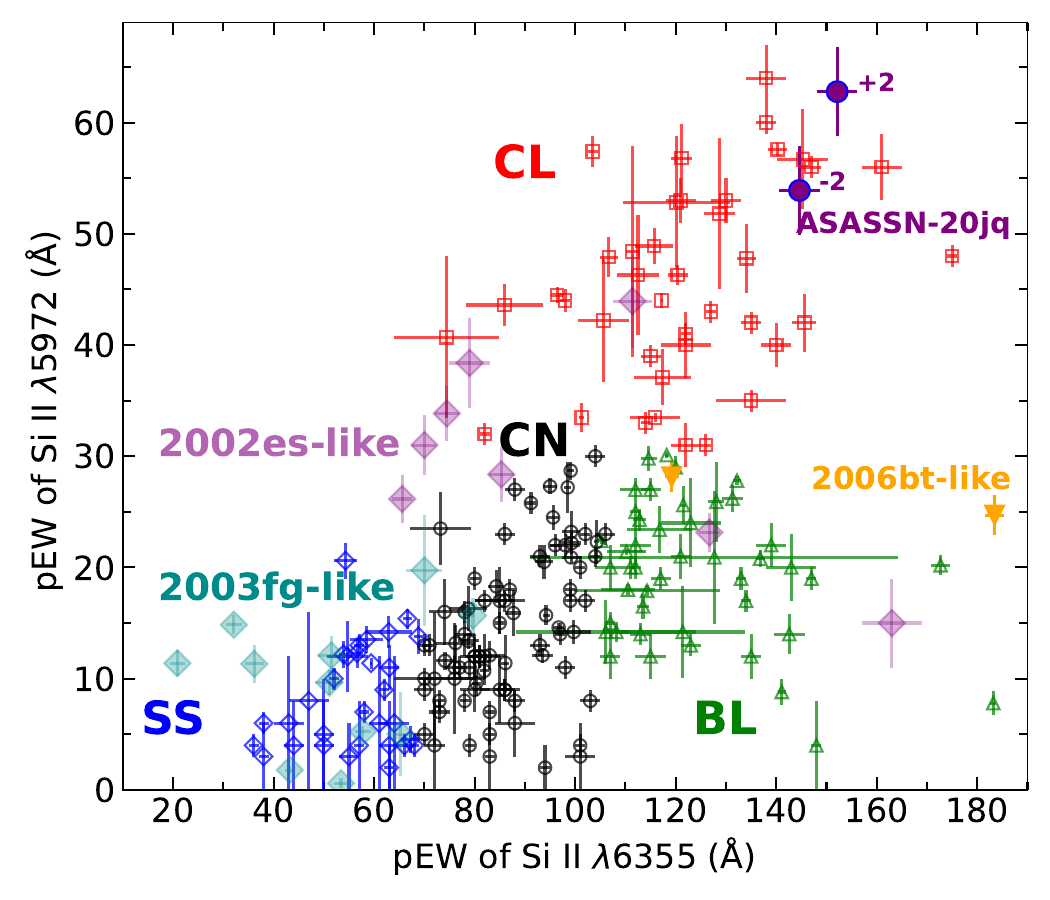}
\caption{Branch diagram comparing the $pEW$ values of the  \ion{Si}{ii} $\ld5972$ and  $\ld6355$ spectral features measured at maximum light. The diagram is populated by the Carnegie Supernova Project  sample of SNe~Ia \citep{Folatelli2013,Morrell2024}. Our measurements from two near-maximum light spectra of \sn\ are plotted in purple, as well as those inferred of nearly a dozen peculiar SNe~Ia with commonalities with SN~2002es (see text).  In addition, we plot the sample of 2003fg-like objects \citep{Ashall2021}. The Branch diagram subtypes labeled are  CN -- core normal, BL -- broad line, CL -- cool, and SS -- shallow silicon.}
\label{fig:branch}
\end{figure}

\begin{figure}[!ht]
	\centering
 \includegraphics[width=\linewidth]{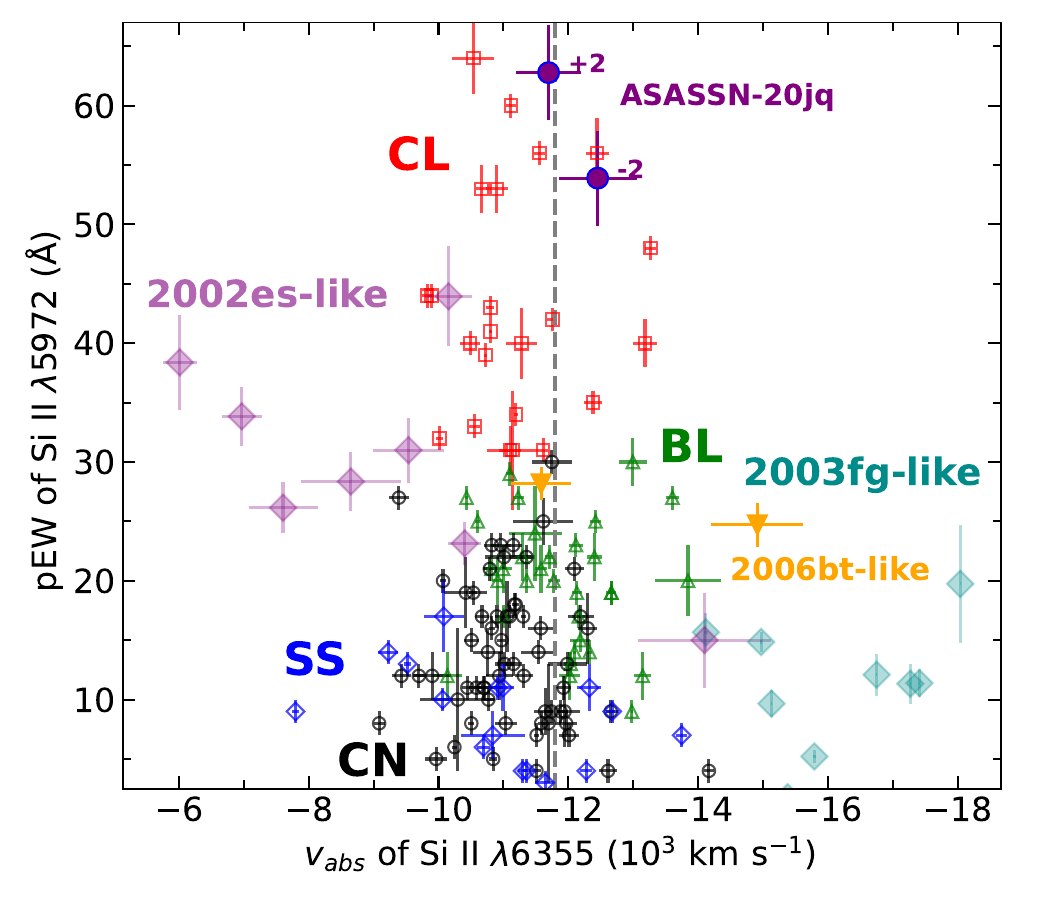}
\includegraphics[width=\linewidth]{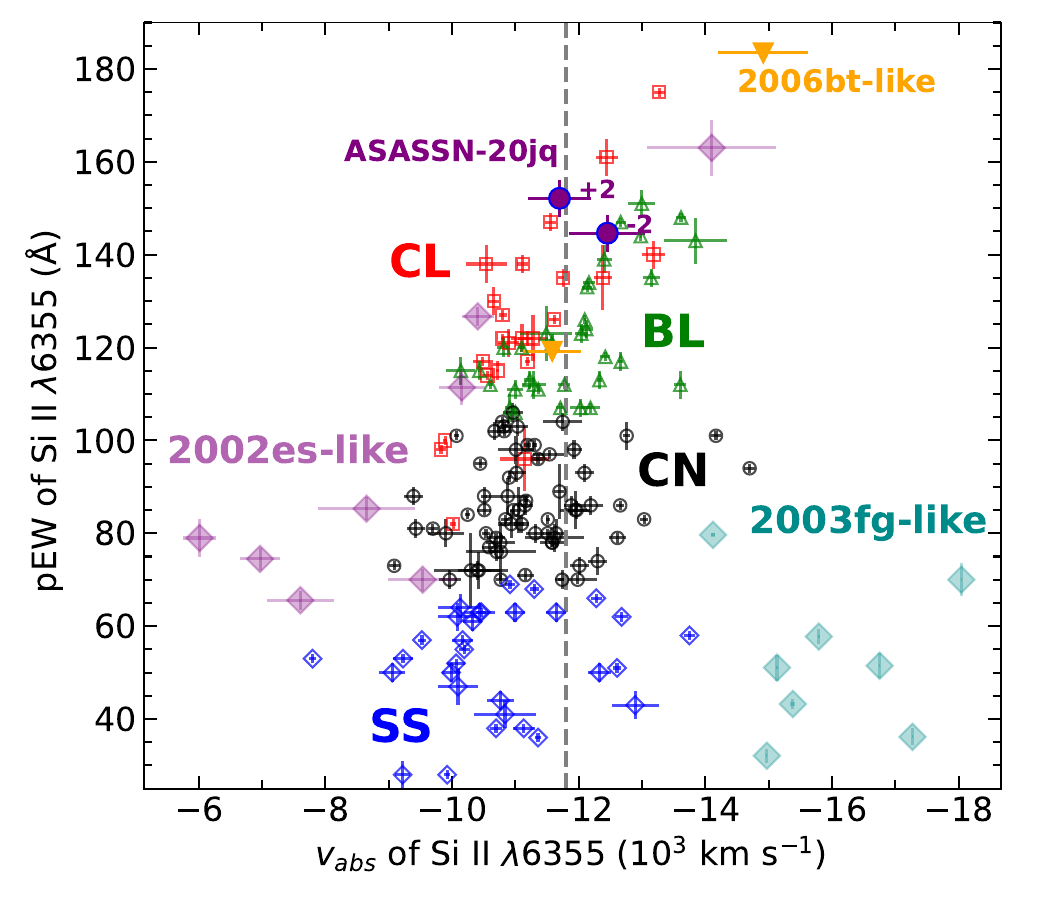}
\caption{ \citet{Wang2009} diagrams comparing the \ion{Si}{ii} $\ldld$5972, 6355 $pEW$ and Doppler velocity ($v(\Siii)_{abs}$) measurements from near maximum spectra of the Carnegie Supernova Project sample \citep{Morrell2024,Ashall2021}. Measurements made from the two nearest to maximum light spectra of \sn\ are plotted  as purple points, as well as those inferred from nearly a dozen peculiar SNe~Ia with commonalities with SN~2002es. Branch types are same as in Fig.~\ref{fig:branch}. The vertical line separates normal and high-velocity SNe~Ia as defined by \citet{Wang2009}. }
\label{fig:Si2pEW_vel}
\end{figure}

\begin{figure*}[!ht]
	\centering
\includegraphics[width=0.75\linewidth]{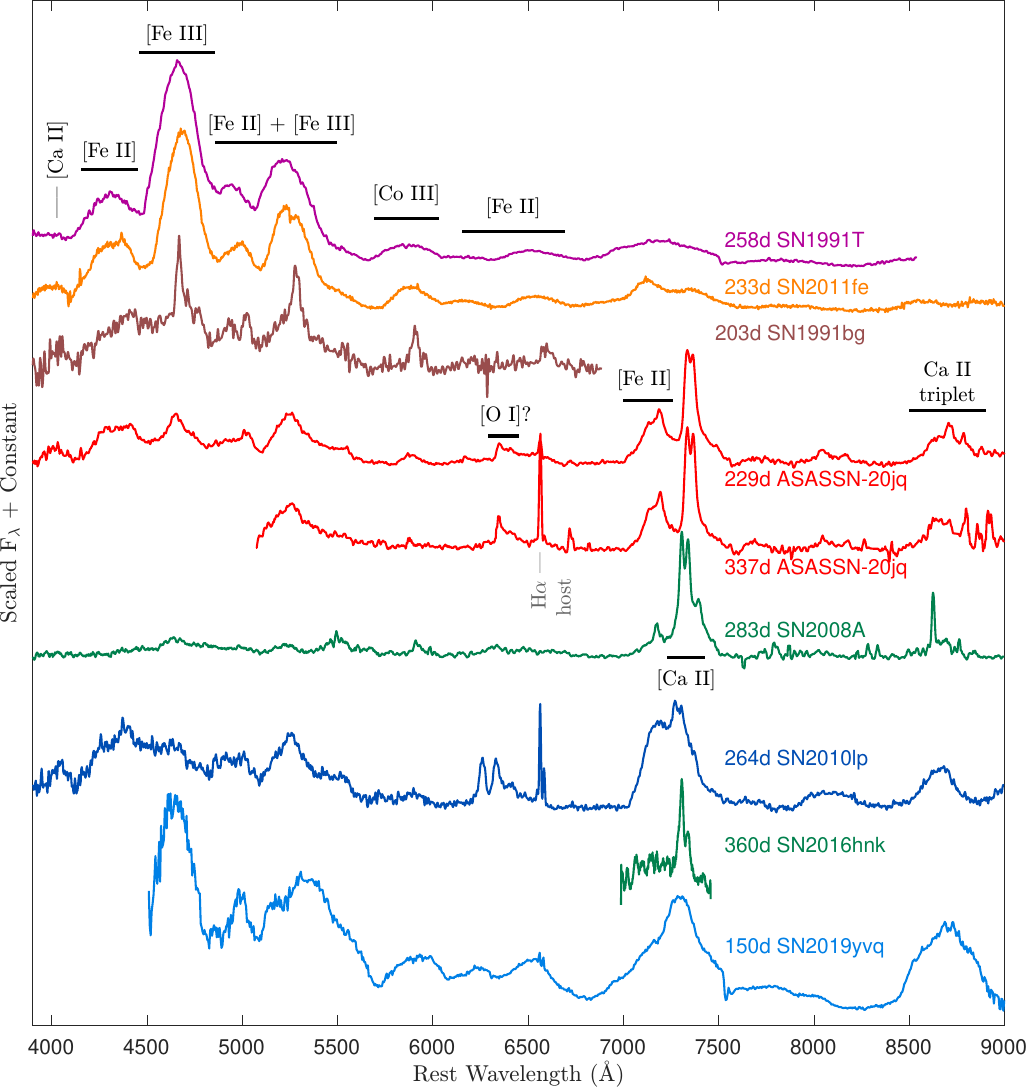}
\caption{Nebular spectra comparison. Spectra of \sn\ obtained at +229~days and +337~days compared with nebular spectra of  SN~1991T \citep{1998AJ....115.1096G}, SN~2011fe \citep{2016ApJ...820...67Z}, the  low-luminosity SN~1991bg \citep{1996MNRAS.283....1T},  the SN~Iax 2008A \citep{McCully2014}, the peculiar SN~2016hnk \citep{Galbany2019}, as well as the 2002es-like 
SN~2010lp \citep{Taubenberger2013}, and SN~2019yvq \citep{Tucker2021}. 
Prominent spectral features are labeled with associated ion following \citet{Wilk2020}.}
\label{fig:spec_comp_neb}
\end{figure*}

We now describe the method used to estimate the $^{56}$Ni mass from the UVOIR light curve of \sn. 
In short, the UVOIR light curve was fit with an energy deposition function associated with the radioactive $\nickel\rightarrow\cobalt\rightarrow\iron$ decay chain. The best-fit model is over-plotted to the UVOIR light curve in the top panel of Fig.~\ref{fig:bol_nimass}. The model fit consists of two free parameters corresponding to the \nickel\ mass ($M_{Ni}$) and the $\gamma$-ray trapping parameter $t_{0\gamma}$. While all the positron kinetic energy from \cobalt\ decay is allowed to be deposited, only $[1- e^{(-t_{0\gamma}^2/t^2)}]$ fraction of $\gamma$-ray energy is trapped in the envelope. At late times, ($>200$ days) when positron energy becomes comparable to the fraction of $\gamma$-ray deposition energy, the model luminosity tends to become degenerate between $M_{Ni}$ and $t_{0\gamma}$. Therefore, in the middle panel of Fig.~\ref{fig:bol_nimass}, we fit the time-weighted integral of the luminosity, which indicates that at all times during the evolution of \sn\ its energy budget can be explained solely by the \nickel\ decay chain \citep{Katz2013}. In other words, there is no evidence from the model fit for any additional sources of energy deposition.  Furthermore, plotted in the bottom panel is  the ratio $t^2L/(\int L t~dt)$ which breaks the late-time degeneracy between $M_{Ni}$ and $t_{0\gamma}$, as it is independent of $M_{Ni}$ \citep{Katz2013}.  This ratio is used to independently and accurately tune the $t_{0\gamma}$ parameter, while the model in the top two panels are used to determine the $M_{Ni}$. The best-fit model corresponds to  $t_{0\gamma}=58\pm5$~days and a low $M_{Ni} = 0.088\pm0.008$~\msol.
Such a low $M_{Ni}$ has been seen only for a few underluminous 1991bg-like SNe \citep{2019MNRAS.483..628S,2020MNRAS.496.4517S}, thereby making \sn\ with one of the lowest $M_{Ni}$ estimate known for SNe~Ia.

\section{Spectroscopic analysis}
\label{sec:spectroscopy}
We now present our analysis of the spectroscopic time-series of \sn.  First the photospheric phase spectra are examined and compared to other similar epoch SNe~Ia covering a range of subtypes, and then   late-phase spectra are considered.

\subsection{Photospheric phase}
\label{sec:earlyspectra}

Figure~\ref{fig:spec_comp_peak}  provides a comparison between the $-$9 days and $-$2~days  spectra of \sn\ to similar phase SNe~Ia representative of a variety of SN~Ia subtypes. This includes pre-maximum spectra of the   1991bg-like SN~1999by \citep{Matheson2008}, SN~1986G \citep{Phillips1987}, SN~1991T  \citep{1995A&A...297..509M}, SN~2006bt \citep{Stritzinger2011},  and  maximum or post-maximum spectra of SN~2011fe  \citep{2013A&A...554A..27P}, the type~Iax SN~2008A \citep{McCully2014}, SN~1991bg \citep{Filippenko1992}, SN~2010lp \citep{Mazzali2022}, SN~2016hnk \citep{Galbany2019}, SN~2002es \citep{Ganeshalingam2012}, and SN~2019yvq \citep{Miller2020}.

At a quick glance the most prominent spectral line features of \sn\ are   consistent with those present in normal SNe~Ia such as SN~2011fe. However, a more detailed  inspection reveals that \sn\ is spectroscopically peculiar, sharing some commonalities with 2002es-like SN~Ia, but also exhibits a number  key  differences.

The  photospheric phase spectra of \sn\  
have a number of conspicuous features which are  labeled in Fig~\ref{fig:spec_comp_peak}. These include the \ion{S}{ii} W-shaped doublet,  the \Siii\ \ld\ld$5972,6355$ doublet,  \Oi\ \ld7773, and the \Caii\ \ld\ld8498, 8542, 8662 near-infrared triplet features. These features are all common to normal SNe~Ia, but typically do not appear  as prevalent as they do in \sn. 
Indeed, similar prevalent absorption profiles are a characteristic of subluminous 1991bg-like and other underluminous 2002es-like objects within the comparison sample.
 From the $-$2~days and $+$2~days spectra of \sn, the    \ion{O}{i} \ld7773 $pEW$ values are measured to be $220\pm5$\,\AA, and $222\pm7$\,\AA. These values are approximately twice as high as those inferred from maximum light spectra of the   subluminous SNe~1991bg ($pEW\approx130$\,\AA) and 2016hnk ($pEW \approx 112\pm4$\,\AA), as measured using the \texttt{Python} package \texttt{spextractor} \citep{Burrow2020}.\footnote{\url{https://github.com/anthonyburrow/spextractor}}

Interestingly, unlike most underluminous SNe~Ia  like SNe~1986G,  SNe~1991bg  and SN~2002es, \sn\ lacks a conspicuous $\sim$4150\,\AA\ \ion{Ti}{ii} absorption feature (see Fig.~\ref{fig:spec_comp_peak}) at maximum light.
However, only the pre-maximum spectrum at $-9$\,days is possibly showing a weak \Tiii\ absorption line.

We now place \sn\ into context with known SNe~Ia spectroscopic relations based on a number of decline-rate and spectroscopic indicators.  These include \ion{Si}{ii} \ld5972 and \ld6355 $pEW$ measurements and the \ion{Si}{ii} \ld6355  Doppler line velocity as measured from the position  of  maximum absorption ($v_{abs}$) measured from spectra obtained within three days of the epoch $B$-band maximum.

The Branch diagram provides a spectroscopic sequence among the SNe~Ia population that reflects the degree of ionization above the photosphere \citep{Nugent1995,2006PASP..118..560B}.  Our Branch diagram  comparing the $pEW$ values of \Siii\ \ld5972 and \Siii\ \ld6355 is  plotted in Fig.~\ref{fig:branch}. 
We measure \Siii\ \ld5972 and \ld6355 $pEW$ values from the $-$2~days  of \sn\ of $54\pm4$\,\AA\ and $145\pm4$\,\AA\, respectively, and from the +2~days spectrum $63\pm4$\,\AA\ and $152\pm4$\,\AA.  
As per the Branch classification scheme, \sn\ falls on the extreme end of `CooL' (CL) type populated by low-luminosity, fast-declining SNe~Ia as well as the so-called transitional SNe~Ia subtype \citep[see, e.g.,][]{Hsiao2015}. 
 \sn\ is an extreme CL type, and with 
 \Siii\ \ld6355  $pEW$ values of $\approx$150\,\AA, also rivals the values exhibited by some of the  ``broadest'' broad line (BL) SNe~Ia in the comparison sample.  

We next investigate other spectroscopic based relations that have similar groupings to the Branch diagram but include information on the velocity of the \Siii\ \ld6355 feature. In Fig.~\ref{fig:Si2pEW_vel}  the near maximum \Siii\  \ld5972  $pEW$   (top) and  \Siii\  \ld6355 $pEW$ (bottom) measurements of \sn\ are plotted versus the  \Siii\ \ld6355 Doppler velocity as inferred from the position of maximum absorption (hereafter $v(\Siii)_{abs}$). The comparison SNe~Ia sample  consists of Carnegie Supernova Project (CSP) objects presented recently by  \citet{Morrell2024}, and augmented  with the CSP 2003fg-like sample \citep{Ashall2021}, SN~2002es \citep{Ganeshalingam2012}, and the  2002es-like objects:  iPTF14atg \citep{Cao2015}, SN~2006bt \citep{Foley2010}, SN~2006ot \citep{Silverman2012}, SN~2010lp \citep{Mazzali2022}, SN~2016ije \citep{Li2023}, SN~2016jhr \citep{Jiang2017},  SN~2019yvq \citep{Miller2020}, SN~2022vqz \citep{Xi2024}, and SN~2022ywc \citep{Srivastav2023}.

Between $-2$ and $+2$ days, \sn\ exhibits  \Siii\  $v_{abs}$ value 
between  $-12\,500\pm600$~\kms to $-11\,700\pm550$~\kms. 
These values place \sn\ near the boundary between the normal and high-velocity SNe~Ia groups, as defined by \citet{Wang2009} at -11\,800~\kms.

\begin{figure}[!ht]
	\centering
	\includegraphics[height=13.1cm]{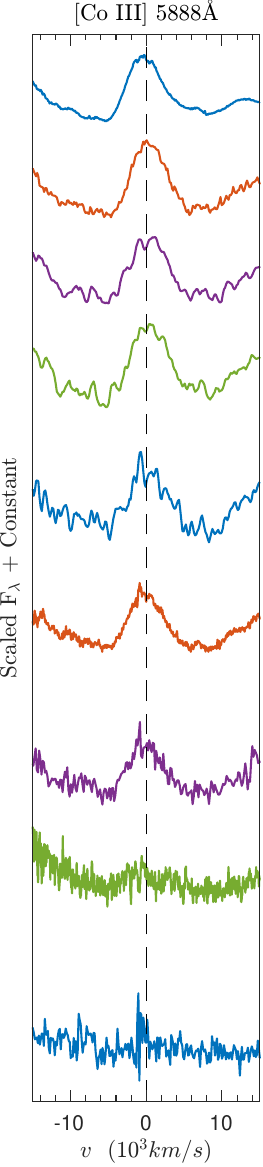}
	\includegraphics[height=13.1cm]{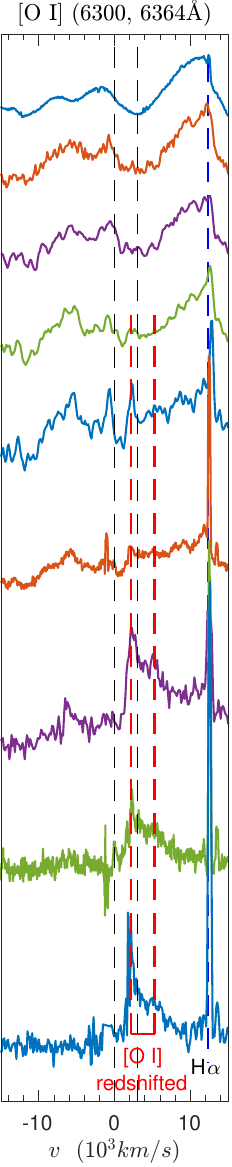}
	\includegraphics[height=13.1cm]{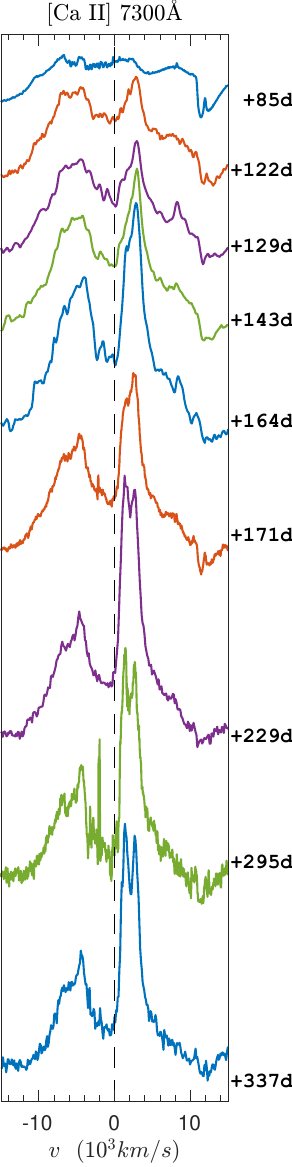}
	\caption{Post-maximum (+85 to +337 days) spectra  of \sn\ plotted in velocity domain with respect to the rest wavelengths, as indicated on top of figure panels, of  [\ion{Co}{iii}] \textit{(left)}, [\Oi] \textit{(middle)} and [\ion{Ca}{ii}] \textit{(right)} lines. The pair of the black dashed lines in the middle panel represents the rest-positions of the [\Oi] doublet components. The pair of red-dashed lines represent the position of the [\Oi] doublet redshifted by $2200\kms$ to best match the observed feature beyond +229\,d. The [\Caii] doublet in right panel shows a systematic redshift of $\approx1700\kms$ throughout the evolution.}
	\label{fig:spec_veldomain}
\end{figure}

\begin{figure}[!ht]
	\centering
\includegraphics[width=0.80\linewidth]{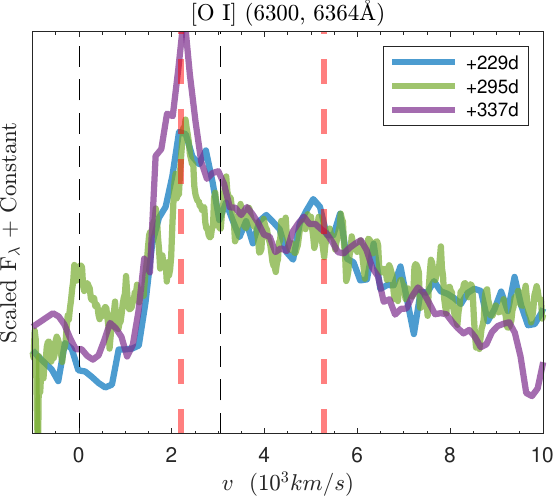}\\
\hspace{-0.8cm}
\includegraphics[width=0.85\linewidth]{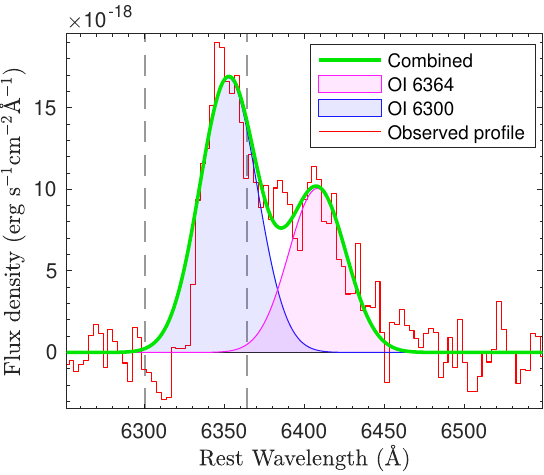}
\caption{Top panel shows the [\Oi] \ldld6300,\,6364 doublet region for the last three nebular spectra of \sn\, are plotted in velocity domain with respect to 6300\,\AA. The pair of black dashed lines represent the rest position of the [\Oi] doublet lines \ldld6300,\,6364, while the dashed red lines represent the [\Oi] doublet redshifted by $\approx2200\kms$. The bottom panel shows the pseudo-continuum subtracted [\Oi] \ldld6300,\,6364 doublet region of the +229d spectrum. A two-component Gaussian profile is fitted with FWHM constrained to be the same. The flux ratio of the doublet components $L_{6300}/L_{6364}$ is estimated to be $\approx1.67$ and the FWHM for each component is $\approx1900\kms$.}
\label{fig:oi_vel_domain_zoom}
\end{figure}

\begin{figure}[!ht]
	\centering
\includegraphics[width=0.95\linewidth]{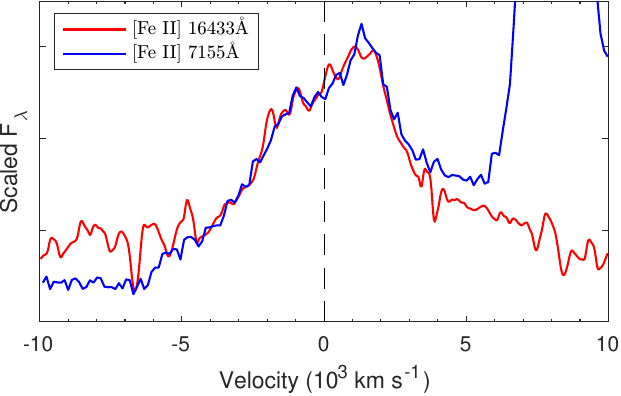}
\caption{Comparison of optical [\Feii] $\lambda$7155 and NIR [\Feii]  $\lambda$16\,433 nebular emission lines plotted in velocity scale. Both lines have identical profiles, being highly asymmetric with a tilted top and peaks redshifted by $\approx 1400$ \kms.}
\label{fig:nebular_feii}
\end{figure}

\subsection{Nebular phase spectroscopy}
\label{sec:nebularspectra}

 Figure~\ref{fig:spec_comp_neb} contains our  nebular phase spectra of \sn\ compared with similar epoch spectra of a  comparison sample. Similar to the comparison sample,  the blue end of \sn's spectrum exhibits several broad feature complexes at the same spectral ranges as the known blends of multiple forbidden [\ion{Fe}{iii}] and [\ion{Fe}{ii}] line transitions observed in SNe~Ia. However, similar to the 2002es-like SN~2010lp, these features in the \sn\ spectra are  more subdued  as compared with the normal, 1991T-like, and even SN~1991bg in terms of line widths and strengths.
 In addition, like normal SNe~Ia,  the $+229$~days spectrum of  \sn\  exhibits forbidden [\ion{Co}{iii}] $\lambda5888$, while at the red end of the  spectrum a conspicuous emission complex  extending roughly between  $\sim 7000 - 7600$\,\AA. 
 In normal SNe~Ia this feature is thought to be a blend of  [\ion{Fe}{ii}] \ld7155 and  [\ion{Ni}{ii}], while for  \sn\ we also see two narrow, partially resolved peaks. Each component is  characterized by a FWHM velocity of $\sim 1200$~\kms\ and are formed by forbidden [\Caii] \ldld7291, 7324 emission. 
 
 Strong [\Caii]  features are  rarely observed at similar epochs in normal SNe~Ia, and when documented typically appear  after $\sim$ +400 days with broad and unresolved doublet components \citep[e.g.,][]{Tucker2022,Kumar2023}. However, the  [\Caii] doublet has  been documented to appear in members of the low-luminosity 2002cx-like and 2002es-like subclasses, as well as a handful of 2003fg-like SNe~Ia. As demonstrated by the Fig.~\ref{fig:spec_comp_neb}, the [\Caii] doublet appear resolved in SN~2016hnk, the 2002es-like SN~2010lp, and the Type~Iax SN~2008A.

Recently, \citet{Siebert2023} reported the presence of a resolved  [\Caii] doublet  with FWHM velocities of 180~\kms\ appearing after $+$240 days in the 2003fg-like SN~2020hvf. However, close inspection of Fig.~\ref{fig:spec_comp_neb} does reveal an inkling of [\Caii] in the $+$258 days spectrum of SN~1991T  \citep[see][for a discussion]{Phillips2024}, while it also appears to contribute to this complex feature in the $+$264 days spectrum of SN~2010lp \citep{Taubenberger2013}.
Finally, with respect to Ca, \sn\  exhibits 
a rather broad emission feature between 8500-9000\,\AA\ formed by the \ion{Ca}{ii} near-IR triplet.  A similar prevalent feature is also present in the comparison spectra of  SNe~2010lp and 2019yvq.

\begin{figure*}[!ht]
\center
\includegraphics[width=0.95\textwidth]{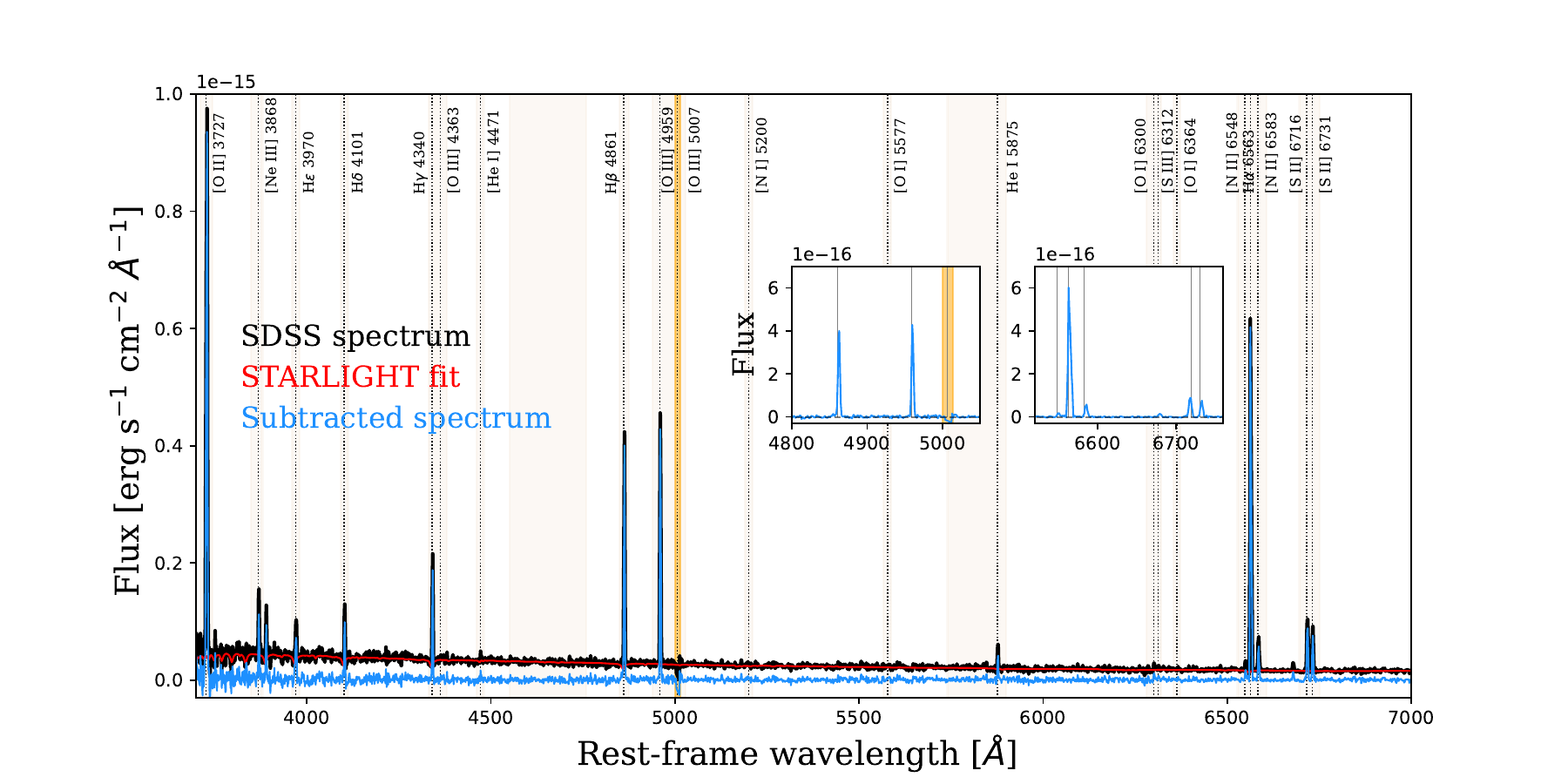}
\caption{The 1D spectrum extracted  (black line) from an SDSS-II spectral observations of \host, obtained from a position in proximity to \sn, and with prominent nebular emission features from the host marked and labeled.
The underlying stellar population was modeled using \texttt{STARLIGHT} (red line) and subtracted from the observed spectrum. The resulting gas-phase spectrum is shown in blue. The most prominent  [\ion{O}{iii}] $\lambda$5007 is saturated and therefore masked by a yellow band.  A single Gaussian function was fit to the nebular lines plotted within the insets. Flux ratios of these lines combined with the calibrations of \citet{Pettini2004} and \citet{Dopita2016} indicate a sub-solar gas-phase metallicity. 
}
\label{fig:galaxy}
\end{figure*}
 
Figure~\ref{fig:spec_veldomain} presents a multi-panel plot of discrete spectral sections in the velocity domain with respect to the rest wavelengths of
[\ion{Co}{iii}] $\lambda5888$, [\Oi] $\lambda\lambda$6300, 6363, and  [\Caii] $\lambda\lambda$7291, 7324.
In the left panel, the [\ion{Co}{iii}]  feature is centered at its rest position. This is juxtaposed with the [\Caii] doublet (right panel) that exhibits a significant redshift of $\sim 1700$\kms\ with respect to its rest wavelength, and this offset remains constant throughout the evolution captured by the time-series  showing in Fig.~\ref{fig:all_spec}. In fact, inspection of the time-series also reveals the [\Caii] features emerge already by $+$85~days. 
Over time the strength of the [\Caii] emission increases relative to the continuum flux, with each component exhibiting an FWHM velocity of $\sim 1200$~\kms. This is reminiscent of some SNe~Iax's, for example,  SNe~2005hk, 2008A and 2010ae \citep{Sahu2008,McCully2014,Stritzinger2014}.

In the middle panel of Fig.~\ref{fig:spec_veldomain}, a prominent and narrow emission line with an extended feature on its red side can be seen (between 2000 to 6000\,\kms) starting from +229\,d. The peak of the feature is close to the $6364$\,\AA, which is the red component of the [\Oi] doublet. However, the lack of any such emission at 6300\,\AA\, poses a problem because at the late nebular phase, in the partially optically thin limit, the 6300\,\AA\, component of the doublet is expected to be equal or stronger in strength compared to the 6364\,\AA\ line (ranges within $1<L_{6300}/L_{6364}<3$ from optically thick to thin limit; e.g., \citealt{1992ApJ...387..309L,2014MNRAS.439.3694J}).
Therefore, we suggest this structure is associated with an [\Oi] doublet (\ldld6300, 6364) redshifted by $\approx2200\kms$, which would imply the prevalent emission peak is the redshifted 6300~\AA\ component. The top panel of Fig.~\ref{fig:oi_vel_domain_zoom}  shows the zoomed region of the tentative  [\Oi] doublet for the last three nebular spectra. In the plot, a pair of vertical solid red lines are included that correspond to the redshifted positions of the [\Oi] doublet. This nicely coincides with the peak and the structure of the emission features. The identification of the redshifted [\Oi] doublet is further corroborated by the similar redshift observed in  the [\Caii] emission doublet. To further validate the [\Oi] line identification, in the bottom panel of Fig.~\ref{fig:oi_vel_domain_zoom} we fit the doublet components and estimate a flux ratio of $L_{6300}/L_{6364}\approx 1.67$, which is well within the expected range of the line transitioning from optically thick to thin limit.
We also considered other possibilities for the origin of this emission feature.
The peak of the emission line coincides with \Siii\, \ld6355, but such narrow emission associated with \Siii\, is not expected at such late phases. We also considered the possibility of \ha\ emission blue-shifted by $\sim10\,000\kms$, though eventually rejected this scenario due to a lack of any similar narrow \hb\ emission feature. 

Beginning by at least +122\,d  \sn\ exhibits a highly asymmetric
tilted-top  [\Feii]~\ld7155 emission line profile that remains throughout the nebular phase. This is identical to the  [\Feii] 1.644~micron line profile of \sn\ as illustrated by Fig.~\ref{fig:nebular_feii}. The blue and red wings of both [\Feii] emission lines are almost symmetrically positioned about the rest wavelengths. However, the asymmetric tilted top of the line profiles makes the peak appear to be shifted red-ward by $\approx 1400 \kms$. As discussed in Sect.~\ref{sec:model} the nebular line profiles provide clues to the explosion physics and viewing angle.

\section{Host-galaxy properties}
\label{sec:host}

The host galaxy of \sn\ is classified as a lenticular, irregular barred spiral (SBm) galaxy, which is the same designation for the Large Magellanic Cloud (LMC). Such galaxies have a disk but lack spiral arms, are of low gas content, and as a result,  typically have low star formation rates (SFRs).

An optical spectrum of the host-galaxy \host\ was obtained by the Sloan Digital Sky Survey (SDSS-II; \citealt{Abazajian2009}). Fortuitously, the location of the instrument fiber was also placed only $3\arcsec$ away from the location of \sn. The extracted 1D spectrum is plotted in Fig.~\ref{fig:galaxy}. The spectrum was fit using the stellar population package \texttt{STARLIGHT} \citep{CidFernandes2005}, providing a best-fit combination of simple stellar population (SSP) models. The resulting fit was removed from the observed spectrum and the resulting pure gas-emission phase spectrum of \host\ is also shown in Fig.~\ref{fig:galaxy}.

The extinction-corrected line flux measurements were used to estimate both the gas-phase metallicity (12 + log(O/H)) and SFR density ($\Sigma$SFR). Unfortunately, the [\ion{O}{iii}] $\lambda$5007 feature is saturated and has been removed from the public spectrum. We therefore turned to the  OH\_N2 calibration from \citet{Pettini2004} and OH\_S2 from \citet{Dopita2016}, providing  metallicity values  of $8.263\pm0.003$ dex and $8.021\pm0.009$ dex, respectively. 
These oxygen abundance values correspond to 0.37 and 0.21 times the solar metallicity value (i.e., $\sim 8.69$~dex, \citealt{Asplund2009}), thereby bracketing the oxygen abundance of the Small Magellanic Cloud (SMC), which is approximately 0.25 times the solar value, or equivalently $\sim 8.03$\,dex \citep[e.g.,][]{Russell1992}. 
This value would predict a significant [\ion{O}{iii}] $\lambda$5007 emission, which is consistent with the feature being completely saturated in the observed SDSS spectrum; see region in Fig.~\ref{fig:galaxy}  highlighted in yellow.
In comparison, this value falls within the bottom 5th percentile of the PISCO sample (\citealt{2018ApJ...855..107G}; updated as of September 2024), which represents the largest collection of SN host galaxies observed using integral field spectroscopy. The most metal-poor SN~Ia host galaxy in this sample has a metallicity of 7.98 dex and is associated with SN~2003du.

We next estimate the SFR which in turn provides an estimate of the SFR density ($\Sigma$SFR) and the specific SFR (hereafter sSFR) of NGC~5002. We measure an H$\alpha$ flux value of (2.47$\pm$0.12) $\times$10$^{-15}$~erg~s$^{-1}$ from the gas-phase stellar-subtracted SDSS-II spectrum. Based on the the \citet{Kennicutt1998} calibration between H$\alpha$ flux and SFR suggests NGC~5002 has a SFR of $(5.85\pm0.02)\times10^{-4} M_\odot$~yr$^{-1}$. Dividing the SFR by the area covered by the SDSS-II fiber in Kpc gives a gas phase SFR density of $\Sigma$SFR = $(1.27\pm0.03)\times10^{-5} M_\odot$ yr$^{-1}$ kpc$^{-2}$. 
This value corresponds to the 50th percentile compared to all PISCO SNe~Ia host galaxies, in other words it has an average sSFR value in relation to the PISCO sample.  
To estimate the  sSFR  we simply divide the SFR estimate above by the stellar mass covered by the fiber, which from the STARLIGHT fit shown in Fig.~\ref{fig:galaxy} indicates $(7.20\pm0.23)\times10^{5} M_\odot$. This implies a  sSFR of $(8.12\pm0.02)\times10^{-10}$ yr$^{-1}$. 
This corresponds to the 50th percentile compared to all PISCO SNe~Ia host galaxies.

Next, from the H$\alpha$ feature in the subtracted SDSS-II spectrum, we measure a high $pEW$ value of $168\pm0.5$\,\AA.
 Using the SSP models from Starburst99 \citep{Leitherer1999}, the $pEW$ value is consistent with ionization of the gas driven  primarily  by a very young stellar population of only a few million years old.
Indeed, the average luminosity weighted age of  $\log_{10} = 7.25\pm1.51$, or roughly  10-15 Myr,  is inferred by our STARLIGHT SSP fitting.
Comparing to the PISCO sample, these two values are in the 1st percentile of the PISCO sample. 

\section{Discussion}
\label{sec:discussion}

\subsection{Comparison of \sn\ with 2002es-like and other similar underluminous SNe~Ia}

\sn\ exhibits many photometric and spectroscopic characteristics that deviate from normal SNe~Ia, but are more consistent with SN~2002es and some of the other members of the 2002es-like subclass. We now highlight some of the similarities among the 2002es-like population, and when appropriate explore the growing observational diversity within the sample. 

\subsubsection{Light curve evolution}

As shown in Fig.~\ref{fig:all_lc}, the $i$-band light curve of \sn\ does not display two distinct peaks; instead, it exhibits a single peak, a characteristic commonly observed among most subluminous 1991bg-like, 2002es-like, and even Iax SNe. Measurement of the peak times for the $B$- and $i$-band light curves yields $t^{i-B}_{max}$ values close to zero. As illustrated in  Fig.~\ref{fig:t_i_B}, this positioning places \sn\ within a relatively sparsely populated area of the parameter space. The figure further shows that, while the expanding sample of 2002es-like and 2006bt-like supernovae tend to have $t^{i-B}_{max}$ values confined to a narrow range between 0 and $+3$ days, their \sbv\ values span a broader range that includes both subluminous 1991bg-like objects and, occasionally, 2003fg-like SN~Ia subclasses.

The intrinsic $B-V$ colors of \sn\ at peak and along the Lira relation are redder than those of normal SNe~Ia, although they are not as red as SN~1991bg (see Fig.~\ref{fig:bv_color}). This trait is typical among 2002es-like objects \citep[][]{Xi2024}. Additionally, as illustrated in Fig.~\ref{fig:dm15sbv}, \sn\ reaches a peak $M_{B}$ that is significantly underluminous relative to its inferred light-curve decline-rate parameter, \dmb, a pattern consistent with the extended 2002es-like comparison sample. Notably, as demonstrated in the right panel of Fig.~\ref{fig:dm15sbv}, even after correcting for light curve shape using the \sbv\ parameter, \sn\ remains a considerable outlier relative to the 2002es-like comparison sample and SN~2016hnk.

In examining the early light curve evolution presented in Sect.~\ref{sec:tfirst}, the flux of \sn\ during the first 1.4\,days of explosion is approximately 2 magnitudes brighter than predicted by the expanding fireball model with a single power-law fit (see Fig.~\ref{fig:tfirst}). A two-day gap between the initial and subsequent photometric epochs limits detailed analysis, yet the presence of early excess emission in the rising light curve of \sn\ agrees with other 2002es-like SNe Ia discovered and followed up shortly after explosion, including iPTF14atg \citep{Cao2015}, iPTF14dpk \citep{Cao2016}, SNe~2016jhr \citep{Jiang2017}, 2019yvq \citep{Miller2020, Burke2021, Tucker2021}, 2022vqz \citep{Xi2024}, and 2022ywc \citep{Srivastav2023}.

In the days following $t_{first}$, each of these SNe~Ia displayed an early flux excess, about two magnitudes or more above what would be anticipated from the expanding fireball model. This initial phase includes a rapid brightening, creating a distinct ``bump'' of varying intensity. For instance, \sn\ and iPTF14atg reached initial absolute magnitudes around $-14.5$ to $-15$ in the optical, while SNe~2016jhr, 2019yvq, and 2020vqz were brighter, around $-16$ to $-16.5$. In the case of the extraordinary SN~2022ywc, the early absolute magnitude reached as high as $-19$. Furthermore, the rise times to maximum brightness across the 2002es-like sample show diversity. iPTF14atg reached peak in approximately 22 days \citep{Cao2016}, iPTF14dpk in about 16 days, SN~2016jhr in 18.9 days, SN~2019yvq in 17.5 days \citep{Burke2021}, \sn\ in roughly 20.4 days,  SN~2022vqz in approximately 18.6 days \citep{Xi2024}, and SN~2022wyc in 20.9 days  \citep{Srivastav2023}. 

We note that the initial outburst radiation field for these SNe are ultraviolet (UV) prevalent, as revealed by Swift in iPTF14atg \citep{Cao2015} and SN~2019yvq \citep{Miller2020, Burke2021, Tucker2021}. However, for \sn\ we lack UV observations to identify such UV excess. Similar UV excesses and blue UV colors have been identified in 2003fg-like SNe~Ia \citep{Hoogendam2024}. In these cases, early UV emission could be a signature of interaction between the expanding SN ejecta and circumstellar material (CSM) \citep[e.g.,][]{Levanon2017, Takashi2023, Maeda2023, Hoogendam2024}.
However, it should also be noted that deep radio observations of several SNe~Ia could never detect early radio emissions to infer any CSM interaction \citep[see e.g.,][]{2016ApJ...821..119C,2020ApJ...890..159L}.
Other scenarios that may explain the rising light curves are
(i.) the interaction between the expanding SN ejecta and a non-degenerate companion \citep{Kasen2010,Maeda2014}, 
(ii.) significant mixing of $^{56}$Ni to the outer layers of the ejecta \citep{Piro2016,Magee2018} or 
(iii.) $^{56}$Ni synthesized in the outer ejecta following a surface ignited double detonation  \citep{Nomoto1982,Woosley1994,Livne1995,Hoeflich1996,Shen2018}, and finally,  
(iv.) the violent merger of two WDs  \citep{Iben1984,Webbink1984,Pakmor2010,Pakmor2012}.

\subsubsection{Spectroscopy}

 The comparison of optical spectra taken around maximum light in Fig.~\ref{fig:spec_comp_peak} reveals a number of commonalities among \sn\ and 2002es-like SNe~Ia. These SNe~Ia exhibit prominent intermediate mass element features of \ion{Si}{ii}~$\lambda\lambda5972,6355$, \ion{O}{i}~$\lambda$7773, and \ion{Ca}{ii} near-infrared (NIR) triplet absorption. The strength of these features in the 2002es-like objects are generally stronger than seen in normal SNe~Ia, suggestive of a  lower ionization state of the underlying emission region \citep{Nugent1995}. We now explore the diversity among key spectral indicators associated with \ion{Si}{ii} and \ion{O}{i} features. We then briefly touch on the  presence or lack thereof of features associated with \ion{Ti}{ii} at $\approx 4150$\,\AA\ and \ion{C}{ii} \ld6580 lines. 

 Referring to the Branch diagram in  Fig.~\ref{fig:branch}, since  2002es-like objects  tend to reach lower peak luminosities than normal SNe~Ia, their underlying ionization state leads to them exhibiting substantial overlap with the CL SN~Ia parameter space, which is primarily  occupied by  transitional and subluminous SNe~Ia. 
 
 The comparison sample of 2002es-like objects also extends beyond this region, moving towards the 2003fg-like subclass and reaching $pEW$ values of \ion{Si}{ii} $\lambda5972$ as low as approximately 20\,\AA. The 2002es-like sample shows minimal, if any, overlap with the SS and CN subclasses of SNe~Ia. Lastly, two objects, SNe 2006bt and 2019yvq, which are not necessarily bona fide members of the 2002es-like group, have broader \ion{Si}{ii} features, placing them among the BL subtype.

Examining the energetics of 2002es-like SNe, as traced by $v(\ion{Si}{II})_{abs}$ (see the Wang diagram in  Fig.~\ref{fig:spec_veldomain}), we find that most of the comparison sample exhibits velocities between $-6,000$ and $-10,000$\kms. However, SN~2006ot and \sn\ reach velocities near the threshold that separates normal from high-velocity SNe~Ia (i.e., $v(\ion{Si}{II})_{abs} \approx -11,500$\kms), while SNe~2006bt and 2019yvr show even higher values, exceeding $-14,000$~\kms.

Focusing on the \ion{O}{i} \ld7773 feature, \sn\ shows a $pEW$ of 220\,\AA, which is nearly twice as strong as in subluminous SNe~Ia and almost four times higher than in CN SNe~Ia \citep{zhao2016}. This value ranks as the highest within the 2002es-like comparison sample, whose measurements include: SN~2002es ($pEW\approx93\pm3$\,\AA), SN~2006bt ($pEW\approx97\pm2$\,\AA), SN~2006ot ($pEW\approx98\pm1$\,\AA), SN~2010lp ($pEW\approx133\pm3$\,\AA), iPTF14atg ($pEW\approx87\pm1$\,\AA), SN~2016ije ($pEW\approx114\pm2$\,\AA), SN~2019yvq ($pEW\approx62\pm5$\,\AA), SN~2022vqz ($pEW\approx70\pm2$\,\AA), and SN~2022ywc ($pEW\approx57\pm2$\,\AA).

Referring to the +3~days spectrum of SN~2002es in Fig.~\ref{fig:spec_comp_peak}, a flux suppression can be seen between 4000-5000\,\AA. A more pronounced absorption  trough  is visible in SN~1991bg (peak $M_B = -16.9$ mag), while a weaker dip is noted in    the transitional SN~1986G (peak $M_B = -17.8$ mag). %
The presence and strength of this trough are temperature-dependent, leading to the formation of \ion{Ti}{ii} \citep{Filippenko1992},  with additional contributions from  \ion{Fe}{ii} and \ion{Mg}{ii} lines \citep[e.g.,][]{Mazzali1997,Doull2011}. Brighter SNe~Ia generally  have higher photospheric temperatures at peak where \ion{Fe}{ii} transitions dominate the blue part of the optical spectrum. In even hotter SNe~Ia, like 1991T-like and 2003fg-like events, \ion{Fe}{iii} features become more prominent due to the higher ionization state.

As seen in Fig.~\ref{fig:spec_comp_peak}, SN~2002es (peak $M_B = -17.8$ mag) and SN~2010lp (peak unknown) display a \ion{Ti}{ii} feature that has been  considered characteristic of the 2002es-like subclass. 
However, in the maximum light spectra of SN~2006bt (peak $M_B = -18.8$ mag), SN~2019yvq (peak $M_B = -18.4$ mag), and \sn\ ($M_B = -17.1$ mag), this trough is absent or very weak. 
The relationship between the feature and luminosity is not straightforward. For instance, SN~2022ywc 
(peak $M_B \sim -19$ mag)
also shows significant suppression in the 4000-5500\,\AA\ range, similar to SN~2016hnk (peak $M_B = -16.7$ mag). %
While the presence of weak \ion{Ti}{ii} features has been suggested in  other 2002es-like SNe, such as PTF10ops (peak $M_B = -17.7$ mag), iPTF14atg (peak $M_B = -17.9$ mag),
SN~2016ije  (peak $M_B = -17.7$ mag), 
and SN~2022vqz  (peak $M_B = -18.1$ mag), we consider the evidence in these cases to be  rather inconclusive.

We conclude our discussion of the photospheric phase spectra by focusing on carbon, which as carbon burning produces oxygen, likely indicates unburned material from the C+O WD progenitor.
 However, only a handful of \ion{C}{ii} lines appear in the optical range, and these are generally quite weak compared to the more prominent spectra features. \ion{C}{ii} features typically emerge in the days following the explosion, with \ion{C}{ii} \ld6580 being the most prominent, often appearing as a notch redward of the \ion{Si}{ii} \ld6533 line. The \ion{C}{ii} \ld6580 feature is observed in the pre-maximum spectra of about 30\% of normal SNe~Ia \citep{Parrent2011,Thomas2011,Folatelli2012,Silverman2012}. It is also prevalent  in the pre-maximum spectra of SN~2006gz \citep{Hicken2007} and in some other 2003fg-like SNe~Ia, such as SN~2009dc \citep{Yamanaka2009}, SN~2007if \citep{Scalzo2010}, SN~2012dn \citep{Taubenberger2019}, LSQ14fmg \citep{Hsiao2020},  PSN J0910+5003 and ASASSN-16ex \citep{Tiwari2023}.

A persistent  and  relatively prevalent \ion{C}{ii} \ld6580 feature has  been identified  in the 2002es-like iPTF14atg with a $v(\ion{C}{ii})_{abs} \approx~-6000$~\kms \citep{Cao2015} and in SN~2016ije with a $v(\ion{C}{ii})_{abs} \approx -4500$~\kms \citep{Li2023}.  \citet{Foley2010} reported evidence of \ion{C}{ii} \ld6580 in the $-4$ and $-3$ days  spectra of SN~2006bt with  $v(\ion{C}{ii})_{abs} \approx -5200$ \kms, and by reaching maximum light the feature  no longer appeared in its spectrum.  Since this velocity is lower than what is typically observed in normal SNe~Ia, 
\citet{Foley2010} suggested that the carbon might result from circumstellar interaction with carbon-rich material.
However, as with normal SNe~Ia, carbon is not universally present in 2002es-like objects, with no evidence of \ion{C}{ii} in the pre-maximum spectra of PTF11ops \citep{Cao2015}, SN~2019yvq \citep{Miller2020}, or SN~2022vqz \citep{Xi2024}. The $-9$ days spectrum of \sn\ shows a notch just redward of the \ion{Si}{ii} \ld6533 emission component, which, if caused by \ion{C}{ii}, would imply an unusually low velocity of $v(\ion{C}{ii})_{abs} \sim -2300$~\kms.
 
We now shift our focus to the nebular optical spectra of \sn\ and other 2002es-like objects, which over time offer an increasingly deeper view into the inner regions of the ejecta.
The nebular spectra of \sn\ resemble those of  SN~1991bg, but with even weaker  Fe-group complexes due to an absence  of [\ion{Fe}{iii}] lines, as shown and labeled in Fig.~\ref{fig:spec_comp_neb}.  
\sn\ closely resembles the 2002es-like SN~2010lp, sharing similarly weak Fe-group element complexes and a prominent double-peaked [\ion{Ca}{ii}] feature. 
Additionally, the nebular spectrum of SN~2010lp displays a complex double-peaked [\ion{O}{i}] \ld\ld6300,6363 emission doublet on top of a broad base.
The emission features of SN~2010lp exhibit a full width at half maximum (FWHM) velocity of approximately 1700\kms. Interestingly, these features are blueshifted and redshifted by 2000\kms and 1800\kms, respectively, relative to the [\ion{O}{i}] \ld6300 rest wavelength, suggesting the presence of an asymmetric oxygen emission region in SN~2010lp \citep{Taubenberger2013}. A similar feature, attributed to [\ion{O}{i}] \ld\ld6300,6363, was observed in the nebular spectrum of iPTF14atg \citep{Kromer2016}, although the low signal-to-noise ratio of the data prevented precise measurements of these features.

As discussed in Sect.~\ref{sec:nebularspectra}  and shown in Fig.~\ref{fig:oi_vel_domain_zoom}, the three nebular spectra of \sn, taken between +229 and +357 days, reveal a feature that likely corresponds to a double-peaked [\ion{O}{i}] \ldld6300,6363 doublet, redshifted by $\approx 2200\kms$, with an FWHM of $\approx 1900\kms$ for each component. Unlike core-collapse SNe, [\Oi] emission is extremely rare to be detected in SNe~Ia.  Based on the analysis of the nebular spectrum of \sn, the redshifted [\ion{Ca}{ii}] and the tentative [\ion{O}{i}] features may be explained by a viewing angle effect combined with an asymmetric explosion (see Section~\ref{sec:model}).

We also considered the possibility of the origin of the redshifted [\Oi] and [\Caii] lines from the leftover material stripped off from a non-degenerate He-rich companion \citep[see][]{2013ApJ...774...37L,2015A&A...577A..39L}. \cite{2013ApJ...774...37L} models suggest that depending on the viewing angle to the axis of SN and companion, these emission lines are shifted by $\sim500 - 2000\,\kms$. \cite{2015A&A...577A..39L} suggested in such a scenario [\Oi] and [\Caii] emission lines of widths of $\sim1000\kms$ would be formed. Although such a scenario might explain the redshifted [\Oi] and [\Caii] lines that we observe in \sn, it is expected that the strength of those lines from stripped-off materials will be much weaker than that we see in \sn. 
For \sn\ the peak intensities are $\sim1.5 - 9.0\times10^{-17}$\,$\rm erg~s^{-1}cm^{-2}\AA^{-1}$ for [\Oi] and [\Caii] lines at +229 days.
\cite{2015A&A...577A..39L} modeled such stripped off emission lines for SNe~2011fe and 2014J, which at the distance of \sn\ are two orders of magnitude weaker than the [\Oi] and [\Caii] line intensities of \sn. Additionally, \cite{2018ApJ...852L...6B} predicted from their non-LTE nebular spectral modeling that lines originating from materials stripped from hydrogen-rich or helium-rich companion would also show \ion{He}{I} lines stronger than  [\Oi] and [\Caii] lines, and in case of \sn\ no \ion{He}{I} lines are detected. Therefore, we disfavor the origin of these lines from stripped-off materials.

We conclude this section by mentioning the host properties of \sn\ in the context of the known characteristics of 2002es-like SN hosts. \citet{White2015} previously noted, based on a limited sample of 2002es-like SNe identified by PTF/iPTF, that these events typically occur in massive early-type, luminous galaxies with minimal star formation. However, as the sample of 2002es-like SNe has grown, exceptions to this trend have emerged. For instance, SN~2016ije was found in a bluer, low-mass star-forming galaxy \citep{Liu2023}, which is more comparable to the host of \sn.

\begin{figure}[!t]
\centering 
\includegraphics[width=1\linewidth]{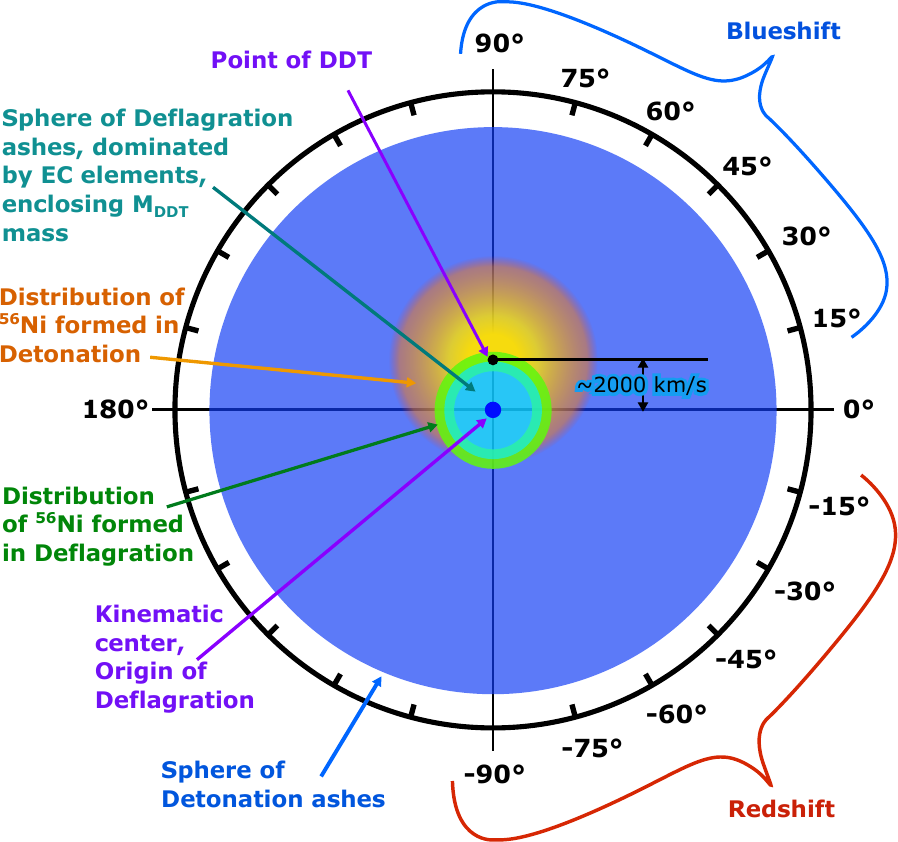}
\caption{Schematic outlining the \citet{Hoeflich2021} off-center DDT explosion model of a $M_{Ch}$-mass C+O WD. Initially, centrally located C+O burning initiates a deflagration flame front, which due to the high central density of the WD, produces an inner, spherical region of electron capture (EC) elements (depicted by filled cyan circle) consisting mostly of stable Fe-group elements.  
The EC  region lays within a spherical shell of remaining deflagration ash consisting mainly of radioactive $^{56}$Ni (green ring). After consuming   0.3 $M_{\odot}$ of the WD and the leading edge of the deflagration flame front reaches $\rho_{tr}$, a $\sim 2000$~\kms\ off-center DDT is triggered (black point). 
As the DDT occurs on a moving background, the detonation burning front requires time to reach and fully consume the remaining C+O ejecta.  As a result,  the detonation burning products are distributed asymmetrically, and as depicted in the schematic by the intensity of yellow which reflects the abundance mass fraction of $^{56}$Ni,  the majority of the $^{56}$Ni produced during the detonation burning phase is distributed in a banana shape region with a ring surrounding the deflagration burning ash and along a positive viewing angle relative to the equator. Depending on the viewing angle, nebular emission lines will appear blue- or red-shifted.}
\label{fig:ddt_schematics}
\end{figure}

\begin{figure*}[!ht]
\centering 
\includegraphics[width=0.9\textwidth]{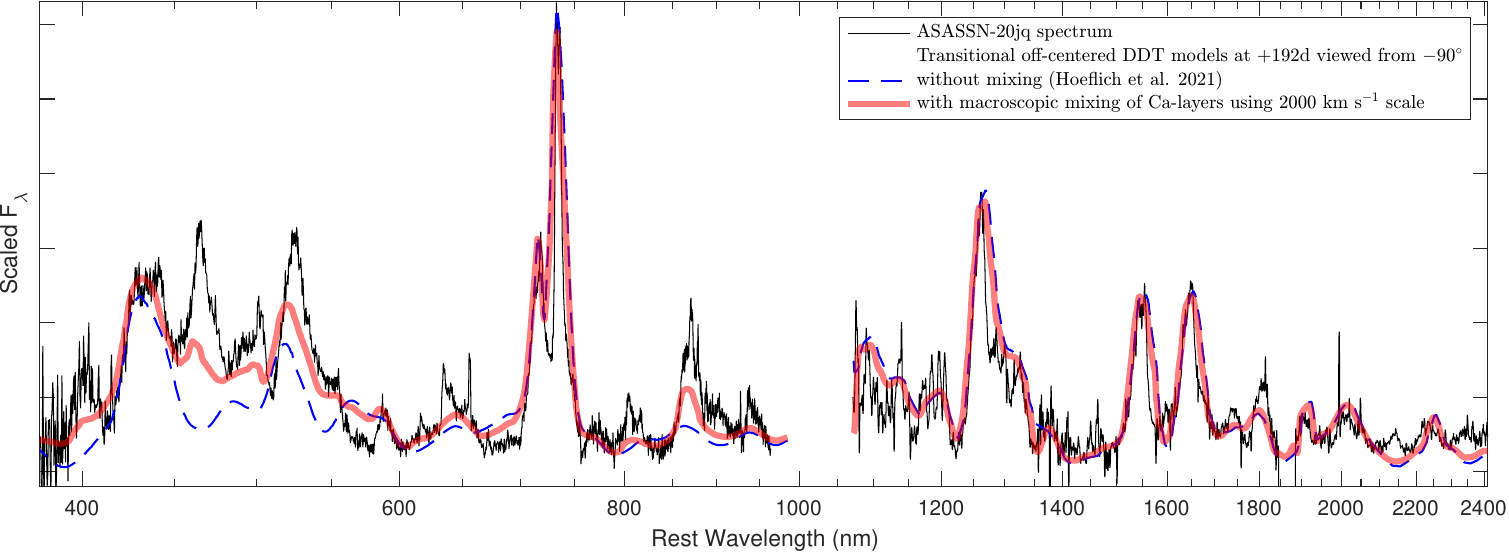}
\caption{Spectral synthesis model of the optical and NIR nebular spectrum. The optical spectrum was obtained on +229 days and the NIR spectrum  on +192~days  \citep{Hoeflich2021}.  Over-plotted in blue is the synthetic spectrum computed from a $M_{Ch}$ C+O WD disrupted following an off-centered delayed detonation explosion. This model  is computed with a viewing angle of $-90^\circ$  as defined by the schematic in Fig.~\ref{fig:ddt_schematics}.
At the blue end of the optical spectrum the model underestimates the flux of several \ion{Fe}{ii}+\ion{Fe}{iii} complexes.  After recomputing the explosion model  with the addition  of a 2000~km~s$^{-1}$ macroscopic mixing  scale of the $^{56}$Ni-Fe into the lower density Ca-layer (see Fig.~\ref{fig:abundance1}), we obtain the synthetic spectrum shown in red.  The mixing model better reproduces the [\ion{Fe}{ii}], [\ion{Fe}{iii}], [\ion{Co}{ii}], and [\ion{Co}{iii}] line complexes  between $\sim 4000-6000$\,\AA\ and $\sim 8500-9000$\,\AA. The observed spectra have been scaled arbitrarily to match the model results. }
\label{fig:theory}
\end{figure*}

\subsection{Extending the \citet{Hoeflich2021} model comparison to optical wavelengths}
\label{sec:model}

\citet{Hoeflich2021} presented a NIR spectrum of \sn\ obtained at +192 days past the epoch of  $B$-band maximum with the W.~M.~Keck Observatory equipped with the Near Infrared Echellette Spectrograph (NIRES), along with a model spectrum computed    with the HYDRA (HYDro RAdiation) transport code  \citep{Hoeflich1990,Hoeflich1995,Hoeflich2003,Hoeflich2009,Hoeflich2017}.
In the following,  the synthetic spectrum at optical wavelengths is compared with the late-phase observed spectrum of \sn. Before doing so, we first present a concise summary of the model details. For a comprehensive description of the numerical methods and model setup, the reader is referred to \citet[][see their Sect.~4 and references therein]{Hoeflich2021}.

\citeauthor{Hoeflich2021} successfully modeled the NIR spectrum of \sn\ with a high-central density WD near accretion-induced collapse, that underwent an off-center DDT, producing (similar to transitional SNe~Ia, e.g., \citealt{Gall2018}) comparable amounts of $^{56}$Ni during the deflagration and detonation phases. 
In this scenario, large-scale asymmetries associated with the
deflagration phase burning products are absent. However, the
off-center DDT triggers a detonation burning phase that produces a
\nickel\ bulge. This asymmetry in the ejecta gives rise to tilted-top
Fe line profiles and prominent [\Caii] emission features (see
below). These unique conditions explain the observed spectral
asymmetries and light-curve peculiarities, distinguishing \sn\ as a
rare SNe~Ia event \citep[see][for details]{Hoeflich2021}. The
  production of comparable amounts of $^{56}$Ni in the deflagration
  and detonation phases is  in contrast to Branch-normal ($^{56}$Ni
  production is dominant in the detonation phase) and subluminous
  SNe~Ia, where \nickel\ production is dominant  in the deflagration phase.
Furthermore, the absence of low-velocity \nickel\ in \sn\ reduces radioactive energy storage in the inner regions, resulting in enhanced energy emission around maximum brightness compared to predictions from Arnett's law \citep{1982ApJ...253..785A}.
 This mechanism in combination with the opacity is linked to the brightness-decline relation for typical SNe~Ia. Consequently, the light curve of \sn\ approaches the radioactive decay curve with a slower post-maximum decline, further distinguishing it from typical SNe~Ia \citep[e.g.,][]{Hoeflich1993,Kasen2006,Hoeflich2017}.

The synthetic spectrum presented in \citet{Hoeflich2021} is the result of a 3-D non-LTE radiative and particle transport calculations based on density and abundance profiles generated from a 2-D  hydrodynamic simulations for the high-density WD 
 disrupted following an off-center DDT.
 The explosion model  is one of a grid of models
 spanning normal-bright to 
 subluminous SNe~Ia \citep{Hoeflich2017}.  
As detailed in \citet{Hoeflich2021},  model 5p02822d40.16 is most aligned with key photospheric phase observables of \sn, including its peak brightness and light-curve decline rate.
 The model consists of a relatively high central density ($\rho_c  = 4 \times 10^9$ g~cm$^{-3}$) C+O $M_{Ch}$-mas  WD evolved from a $5~M_\odot$ main sequence star with solar metallicity. This high central density is a requirement to produce the flatter profile with a tilted top of [\Feii] lines, as seen in Fig.~\ref{fig:nebular_feii}.
  While the
  environment around \sn\ indicates low metallicities (similar to those
  of the SMC), the models presented here did not account for this. The
  metallicity effect on the nuclear burning should be small, since the
  density dominates the electron capture rates. However, the low
  metallicity would affect the cooling and hence the accretion rate in
  single progenitor and secular merger scenarios. The high densities
  we find require low accretion rates, which is compatible with lower
  metallicity, but this was not directly accounted for in our modeling.
 
 Figure~\ref{fig:ddt_schematics} provides a schematic of the adopted explosion model. Initially, a thermonuclear runaway occurs within the central region of the WD with nuclear burning occurring within a  subsonically propagating deflagration flame front. As the central region of the WD is consumed, the WD's core expands with  a velocity of $\sim 2000-3000$~\kms, leaving within its ashes  an inner region consisting mostly of stable electron capture (EC) elements (e.g., $^{54}$Fe, $^{54}$Co and $^{58}$Ni), surrounded by a spherical shell of mostly radioactive $^{56}$Ni. Upon burning $M_{DDT} = 0.3 M_{\odot}$ of the WD  and the leading edge of the deflagration flame front reaches the transition density of $\rho_{tr} = 1.6\times10^7$ g~cm$^{-3}$, an off-center detonation  is launched at a single point located along the north pole extending from  the center of the WD.  As the off-center DDT occurs on an already  expanding background, it  takes time for the detonation burning front to propagate to the opposite side of the WD. When the burning front does reaches the opposite side of the WD, burning occurs under lower densities (as density decreases cubically over time) and hence at lower temperatures. Consequently, an asymmetric distribution of all burning products including Ca and $^{56}$Ni is produced during the detonation burning phase.  As indicated in Fig.~\ref{fig:ddt_schematics}, the asymmetric distribution  of $^{56}$Ni leads to a banana shape ring distributed between  $\sim 0^\circ$ to $180^\circ$ azimuthally  around the outer deflagration ash. The ring becomes thicker with increasing brightness and decreasing central density of the WD. In addition to $^{56}$Ni, the  distribution of $^{40}$Ca  produced during the detonation phase is also asymmetric, and is distributed following  a similar banana-shape with a ring distribution as the $^{56}$Ni.
  This naturally leads to the viewing angle effects on the line profiles as demonstrated in  \citet[][see their Fig.~3]{Hoeflich2021}.

 Fig.~\ref{fig:theory} compares the optical/NIR nebular spectrum of \sn\ along with the synthetic spectrum of model 5p02822d40.1. 
With an $\approx 40$ day difference in phase,  the optical/NIR nebular spectra are not exactly  matched in phase, however, as the spectral features evolve slowly (see Fig.~\ref{fig:spec_comp_neb}) and the synthetic spectrum corresponds to $\sim +$205~days,  the comparison can be considered  robust. 
The model generally matches well with the optical spectral range, particularly with the prominent [\ion{Ca}{ii}] feature and most of the major line complexes. 
The prevalent  [\ion{Ca}{ii}] feature in the synthetic spectrum is produced by $^{40}$Ca created during the detonation burning phase.
The [\ion{Ca}{ii}]  emission is  a consequence  of the significant overlap between the abundance profiles of $^{40}$Ca  and $^{56}$Ni
(see Fig.~\ref{fig:abundance1}).
Furthermore, the [\ion{Ca}{ii}]  feature is  double peaked and redshifted  as 
natural consequence of a $-90^o$ viewing angle  \citep[see also][their Fig. 4 inset]{Hoeflich2021}. 

Turning our attention to bluer wavelengths,  the synthetic spectrum does show flux deficiencies relative to the observed spectrum at the positions of several key Fe-group element complexes. This includes  line complexes associated with higher ionization lines such as: [\ion{Co}{ii}] $\lambda$4154, [\ion{Co}{ii}] $\lambda$4624, [\ion{Fe}{iii}] $\lambda$4659,
[\ion{Co}{iii}] $\lambda$5890,
 [\ion{Fe}{iii}] $\lambda$5272,
 [\ion{Co}{iii}] $\lambda\lambda$6129,6197, 
  [\ion{Co}{ii}] $\lambda$8123, and
 [\ion{Fe}{ii}] $\lambda$8619. 

We exclude  both changes in the inclination angle towards a more equatorial directions, and the $^{56}$Ni abundance as the culprit of the discrepancy given the inherent degradation to the  line profiles including both the conspicuous  optical  [\ion{Ca}{ii}] $\lambda\lambda$7291,7324 and NIR [\ion{Fe}{ii}] $\lambda$1.644 $\mu$m features.  
Other possibilities could be related  to physical 3-D effects not included in the  model. This could take the form of either microscopic and/or macroscopic mixing driven by velocity fields produced during the deflagration burning phase. Microscopic mixing is known to enhance the Si and S lines in the NIR for both normal-bright and subluminous SNe~Ia, but does not  boost the optical Fe-group complexes \citep[see][]{Hoeflich2002,Diamond2018}. 
On the other hand, macroscopic mixing of the $^{56}$Ni to lower-density layers effectively reduces  recombination, which in turn produces a higher-ionization state and thus increased emission from high ionization lines  such as those associated with [\ion{Fe}{iii}] and [\ion{Co}{iii}], and the permitted \ion{Ca}{ii} NIR triplet. 

To this end, macroscopic, radial mixing was implemented into the benchmark model  which alters the final abundance profiles. 
This was  accomplished by including a velocity scale growing from 20\% of the local expansion velocity to a maximum value of 2000~\kms,  which is the expected typical value of  Rayleigh-Taylor instability driven large plumes of burnt material and corresponding downdrafts of unburnt material ubiquitous to deflagration models \citep{Khokhlov2001,Gamezo2003}. The outcome of the inclusion of macroscopic mixing  is revealed by  Fig.~\ref{fig:abundance1}  which contains  the original model and the mixed model abundance profiles.
 The inclusion of macroscopic mixing effectively leads to an increased overlap (in velocity space) between the $^{56}$Ni-rich region and the Ca region. 
 As can be seen in  Fig.~\ref{fig:theory}, the mixed model shows much better agreement with the previously noted Fe-group element complexes and the \ion{Ca}{ii} NIR triplet.
Further comparison between the   observed and synthetic spectrum reveals the model underestimates the flux of the purported [\ion{O}{i}] feature as well as another small  feature around 8000\,\AA\  likely formed by  \ion{Si}{ii} \ld\ld8044, 8243 and straddled between the [\ion{Ca}{ii}] and \ion{Ca}{ii} NIR triplet.    The [\ion{O}{i}] could possibly be strengthened by additional mixing of oxygen into deeper layers.

Turning to the nebular Fe-group line profiles associated with the forbidden lines  of  [\ion{Co}{iii}] $\lambda$5888, [\ion{Fe}{ii}] $\lambda$7155 and [\ion{Fe}{ii}] 1.644~$\mu$m, and the intermediate-mass elements (IMEs) of forbidden  [\ion{O}{i}] \ldld6300,6364, and [\ion{Ca}{ii}] \ldld7291,7324, inspection of Figs.~\ref{fig:spec_veldomain} and \ref{fig:nebular_feii} reveals that the Fe-group lines  show no velocity shifts relative to their rest wavelengths, while the IMEs lines of  [\ion{O}{i}] and [\ion{Ca}{ii}]  show red-shifts  on the order of  2200 \kms and 1700 \kms, respectively. The [\ion{O}{i}] lines are narrow, the FWHM of the Gaussian fit of Fig.~\ref{fig:oi_vel_domain_zoom} is only $\sim 1900$~\kms, implying that the excited oxygen is confined to a relatively narrow region of velocity space.  In relation to the off-center DDT model, as $\sim$50\%
of the synthesized \nickel\  is produced during the deflagration  burning phase, it is distributed 
centrally in a roughly spherical configuration. On the other hand,   the O and Ca features come from  elements synthesized during the detonation burning phase and hence are located in the inner edge of the material above the point of the DDT as defined in Fig.~\ref{fig:ddt_schematics}. The narrow velocity extent is due to the fact that only excited oxygen emits and there can be more oxygen in the ejecta than just that seen in the optical spectrum.

\section{Summary}
\label{sec:summary}

We have presented a detailed analysis and comparison of optical photometry and spectroscopy of the  SN~Ia \sn\ located in a metal poor galaxy with a very young stellar population, and average star formation rates.  
\sn\ is a low-luminosity object peaking at an absolute $B$-band magnitude $M_B = -17.1$ mag.
The light-curve shape parameters of $\dmb=1.35\pm0.09$ mag and $\sbv\gtrapprox0.82$ are well within the range of normal SNe~Ia. 
Combined with the inferred peak absolute magnitude  makes \sn\ a prominent outlier in both the luminosity-width and luminosity-color-stretch relations, appearing $\sim 2.5$\,mag fainter than implied by these calibration relations, which has not been seen for any other SNe~Ia. 
The $i$-band post-maximum light curve of \sn\ does not show a secondary maximum, but rather a flatter evolution. This is characteristic of  1991bg-like SNe~Ia as opposed to normal SNe~Ia  exhibiting distinct double peaks.

We estimated the explosion epoch of \sn\, to be $-20.4$\,days before the $B$-band maximum, by assuming a power-law early-time rising light-curve while taking into account the first detection and non-detection limits. This rise-time to maximum is within the observed diversity of 2002es-like SNe. \sn\ is estimated to have a \nickel\ mass of $0.088\pm0.008$\msun, making it one of the lowest \nickel\ mass estimate for SNe~Ia. Only a few underluminous  1991bg-like SNe are known to have a comparable \nickel\ mass.

The photospheric phase spectra of \sn\ are mostly similar to 1991bg-like and under-luminous 2002es-like objects. However, it lacks a strong $\sim4150$\,\AA\ \Tiii\ feature at maximum light, which is characteristic of most under-luminous objects. \sn\ shows unusually strong \Siii\ \ldld5972, 6355 absorption lines and, as evident from the Branch diagram, the \Siii\ line pEWs are stronger than most normal or under-luminous objects. In addition, \sn\ also shows unusually strong \Oi\ \ld7773 and \Caii\ NIR triplet lines which are much stronger (pEW $\gtrsim2$ times) than those seen in normal or under-luminous objects.

Nebular phase optical spectra of \sn\ are dominated by [\Feii], [\ion{Fe}{iii}] and [\ion{Co}{iii}] lines, albeit much weaker than in normal luminosity SNe~Ia, but similar to under-luminous SNe. However, the [\Feii] \ld7155 line being an exception, which is unusually stronger than that seen in normal, luminous or underluminous SNe (except the 2002es-like SN~2010lp). We also identify identical tilted-top line profiles for both optical and NIR [\Feii]  \ld7155 and \ld16433 lines, which can be attributed to the viewing angle effect of off-center DDT explosion in the SN. Due to the tilted top of the [\Feii] lines, there is an apparent redshift of the peak position by $\approx1400$\kms. However, the overall structure including the red- and blue-wings is symmetrically positioned around the rest wavelength, indicating no shift in the bulk of the [\Feii] emission.
Another striking feature in the nebular phase spectra is the strong and partially resolved  [\Caii]  \ldld7291, 7324 doublet, which can be identified from +122d onward. Except for a few SN-Iax, such strong and resolved [\Caii] doublet lines are never seen in SNe~Ia. In rare cases, strong [\Caii] has been identified in normal SNe~Ia only at much later phases ($\sim400$\,days), but those are always broad with unresolved doublet components. Additionally, this entire [\Caii] doublet profile is redshifted by $\approx1700$\kms.
At $\sim$6350\,\AA, all three nebular spectra of \sn\ show an emission feature with extended red wing. Although limited by the low SNR of our spectra,  we identify these features to be most likely the [\Oi] \ldld6300, 6364 doublet redshifted by $\approx2200$\kms.  The 6300\,\AA\ component of the doublet corresponds to the prominent peak of the emission complex, while the pEW of the 6364\,\AA\ component is weaker by a factor of 1.67, and is dominated by the spectral noise. Detection of [\Oi] doublet in thermonuclear SNe is extremely rare, having only been conclusively detected in SN~2010lp \citep{Taubenberger2013} and SN~2022pul \citep{2024ApJ...966..135K}. 

Interestingly, the forbidden emission lines from intermediate mass elements, viz. [\Caii] and [\Oi], produced during the detonation burning phase, show a redshift of roughly 2000\kms. This is consistent with the off-center DDT scenario for \sn, where the point of detonation is shifted by a few times $10^3$\kms\ in velocity space with respect to the kinematic center of the explosion. On the other hand, the iron group emission lines of [\Feii], [\ion{Fe}{iii}] and [\ion{Co}{iii}] show no discernible shift, as around half of \nickel\ is produced during the initial deflagration phase and is centrally distributed. Using the off-center DDT model we synthesized the optical and NIR nebular spectra simultaneously and also introduced macroscopic mixing of the \nickel-rich region with the low-density Ca-rich layers, which is in good agreement with the observed spectra.
Within $M_{Ch}$ explosion models, objects like \sn\ are
  expected to be rare compared to the population of Branch-normal and
  subluminous SNe~Ia. This rarity arises from the specific
  requirements: a high central-density WD that undergoes an off-center
  DDT explosion, producing nearly equal amounts of \nickel\ during
  both the deflagration and detonation burning phases.

\sn\ shows several commonalities with 2002es-like objects, beyond just a faint peak absolute magnitude, normal SNe~Ia like \dmb\ decline rate, and lack of secondary \textit{i}-band maximum. Spectroscopically, \sn\ also shows a number of similarities with 2002es-like objects, including the early appearance of detectable [\Caii] at +122\,days, and strong \Siii\ \ld6355 and \Oi\ absorption features.
However, 2002es-like objects are a very diverse group of SNe~Ia without a well-defined parameter space, and \sn\ adds to the diversity of this class. \sn\ also lacks a number of features that are seen in some (but not all) 2002es-like SNe, including the lack of a \Tiii\ $\sim 4150$\,\AA\ feature around peak, peak-magnitudes fainter than most well-defined 2002es-like objects, and a lack of un-burnt carbon in the spectra. 
However, other spectral characteristics including the \Siiii\ line velocities are aligned with normal  or overluminous 1991T-like SNe~Ia.

\begin{acknowledgement}
S.B. and M.D.S. are funded by the Independent Research Fund Denmark (IRFD) via Project 2 grant 10.46540/2032-00022B.
M.D.S. and E.J. are supported by the Aarhus University Nova grant\# AUFF-E-2023-9-28.
L.G. acknowledges financial support from AGAUR, CSIC, MCIN and AEI 10.13039/501100011033 under projects PID2023-151307NB-I00, PIE 20215AT016, CEX2020-001058-M, ILINK23001, COOPB2304, and 2021-SGR-01270.

NUTS use of the NOT is funded in part by the Instrument center for Danish Astrophysics (IDA). 
This paper make use of observations from LBT, which is an international collaboration among institutions in the United States, Italy and Germany. LBT Corporation partners are: The University of Arizona on behalf of the Arizona Board of Regents; Istituto Nazionale di Astrofisica, Italy; LBT Beteiligungsgesellschaft,
Germany, representing the Max-Planck Society, The Leibniz Institute
for Astrophysics Potsdam, and Heidelberg University; The Ohio State
University, representing OSU, University of Notre Dame, University
of Minnesota and University of Virginia.
E.B., C.A.,  and  P.H. acknowledge support from NASA grants JWST-GO-02114,
JWST-GO-02122, JWST-GO-04522, JWST-GO-04217, JWST-GO-04436,
JWST-GO-03626, JWST-GO-05057, JWST-GO-05290, JWST-GO-06023,
JWST-GO-06677, JWST-GO-06213, JWST-GO-06583. Support for
programs \#2114, \#2122, \#3626, \#4217, \#4436, \#4522,  \#5057,
\#6023, \#6213, \#6583, and \#6677
were provided by NASA through a grant from the Space Telescope Science
Institute, which is operated by the Association of Universities for Research in
Astronomy, Inc., under NASA contract NAS 5-03127.
P.H. acknowledge support for the simulations by the NSF-grants AST-1715133, and AST-2306395.
This material is based upon work supported by the National Science Foundation Graduate Research Fellowship Program under Grant Nos. 1842402 and 2236415. Any opinions, findings, conclusions, or recommendations expressed in this material are those of the author(s) and do not necessarily reflect the views of the National Science Foundation.
A. F. acknowledges the support by the State of Hesse within the Research Cluster ELEMENTS (Project ID 500/10.006)
AR acknowledges financial support from the GRAWITA Large Program Grant (PI P. D’Avanzo).
AR, NER and AP acknowledge support from the PRIN-INAF 2022 ``Shedding light on the nature of gap transients: from the observations to the models''.
JTH is supported by NASA grant 80NSSC23K1431.
This work is based on observations made with the Gran Telescopio Canarias (GTC), installed in the Spanish Observatorio del Roque de los Muchachos of the Instituto de Astrofísica de Canarias, in the island of La Palma.
A.C. has been supported by ANID Millennium Institute of Astrophysics (MAS) under grant ICN12\_009.
CPG acknowledges financial support from the Secretary of Universities and Research (Government of Catalonia) and by the Horizon 2020 Research and Innovation Programme of the European Union under the Marie Sk\l{}odowska-Curie and the Beatriu de Pin\'os 2021 BP 00168 programme, from the Spanish Ministerio de Ciencia e Innovaci\'on (MCIN) and the Agencia Estatal de Investigaci\'on (AEI) 10.13039/501100011033 under the PID2023-151307NB-I00 SNNEXT project, from Centro Superior de Investigaciones Cient\'ificas (CSIC) under the PIE project 20215AT016 and the program Unidad de Excelencia Mar\'ia de Maeztu CEX2020-001058-M, and from the Departament de Recerca i Universitats de la Generalitat de Catalunya through the 2021-SGR-01270 grant.
\end{acknowledgement}
\clearpage

\bibliographystyle{aa}
\bibliography{ms}

\begin{appendix}
\onecolumn
\FloatBarrier
\section{Optical Broad-band Photometry of \sn}
\label{sec:optphotometrytable}

Table~\ref{tab:photsn} lists the optical broad-band photometry of \sn.

{%
\centering
\fontsize{2.3mm}{3.3mm}\selectfont
\begin{longtable}{c c r c c c c c c l}
\caption{Photometry of \sn.}\label{tab:photsn}\\
\hline\hline
UT Date &        
JD $-$& 
Phase\tablefootmark{a} &  
$B$ &
$V$ &                  
$g$ &                  
$r$ &                  
$i$ &                  
$z$ & 
Telescope\tablefootmark{b} \\ 
&  
2,459,000    &   
(days) &              
(mag) &              
(mag) &             
(mag) &              
(mag) &             
(mag) &         
(mag)     &      
/ Inst.\\
\hline
\endfirsthead
\caption{continued.}\\
\hline\hline
UT Date &        
JD $-$& Phase &  
$B$ &
$V$ &                  
$g$ &                  
$r$ &                  
$i$ &                  
$z$ & 
Telescope \\ &  
2,459,000    &   
(days) &              
(mag) &              
(mag) &             
(mag) &              
(mag) &             
(mag) &         
(mag)     &      
/ Inst. \\
\hline
\endhead
\hline
\endfoot
2020-08-04.14  &   65.64   &   $-$20.1    &    ---               &      ---            &   >16.90            &       ---           &    ---              &         ---          &          ASAS-SN \\
2020-08-04.29  &   65.79   &   $-$19.9    &    ---               &      ---            &         ---         & >18.46\tablefootmark{c}              &    ---              &         ---          &          ATLAS-O \\ 
2020-08-05.25  &   66.75   &   $-$19.0    &    ---               &      ---            &         ---         &       ---           &    ---              &    17.41 $\pm$ 0.030\tablefootmark{d} &        PanSTARRS \\  
2020-08-08.13  &   69.63   &   $-$16.1    &    ---               &      ---            &   16.800 $\pm$ 0.500&       ---           &    ---              &         ---          &          ASAS-SN \\
2020-08-09.16  &    70.66  &    $-$15.1   &   17.336 $\pm$ 0.055 &                ---  &  17.073 $\pm$ 0.040 &  16.533 $\pm$ 0.016 &  16.539 $\pm$ 0.040 &                ---   &            PO-NM \\
2020-08-09.89  &    71.39  &    $-$14.3   &   17.173 $\pm$ 0.077 &  16.602 $\pm$ 0.040 &  16.944 $\pm$ 0.038 &  16.497 $\pm$ 0.027 &                ---  &                ---   &       Asiago-0m9 \\
2020-08-11.13  &    72.63  &    $-$13.1   &   16.708 $\pm$ 0.046 &                ---  &  16.483 $\pm$ 0.051 &  16.081 $\pm$ 0.039 &  16.146 $\pm$ 0.046 &                ---   &            PO-NM \\
2020-08-12.13  &    73.63  &    $-$12.1   &   16.444 $\pm$ 0.051 &                ---  &  16.153 $\pm$ 0.038 &  15.683 $\pm$ 0.028 &  15.908 $\pm$ 0.045 &                ---   &            PO-NM \\
\end{longtable}
\tablefoot{\tablefoottext{a}{Rest-frame days relative to the epoch of $B$-band maximum, i.e., \PeakEpoch.}
\tablefoottext{b}{The abbreviations for the telescope/instrument are: ASASSN - ASAS-SN quadruple 14cm telescopes; {ATLAS-O - ATLAS survey telescope's orange filter observation; PanSTARRS - Pan-STARRS survey telescopes; ZTF - ZTF survey telescopes; Asiago-0m9 - 0.9 m Schmidt telescope at Asiago; PO-NM - Post Observatory 0.6\,m telescope; PO-NM-0m8 - Post Observatory 0.8\,m telescope;} LCOGT-2m - Las Cumbres Observatory 2\,m telescope network. Data observed within 5\,hr are represented under a single-epoch observation.}
\tablefoottext{c}{The reported magnitude is in ATLAS-Orange filter.}
\tablefoottext{d}{This $z$-band photometry is in the PanSTARRS filter system.}
}

}

\FloatBarrier

\section{Supplementary material of the nebular model}
\label{sec:nebularmodlelines}

\begin{figure*}[!h]
\centering
\includegraphics[width=0.9\textwidth]{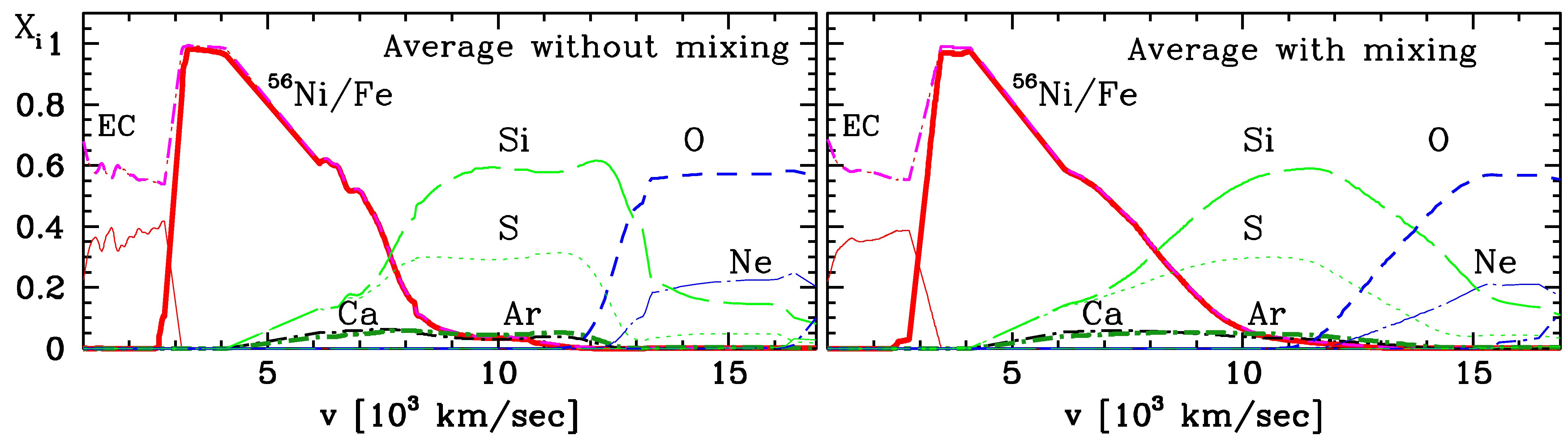}
\caption{The angle averaged composition of the multi-dimensional reference model  5p02822d40.16 
 with macroscopic mixing up to 2000~\kms\ (right), and  without (left).}
\label{fig:abundance1}
\end{figure*}

Plotted in Fig.~\ref{fig:abundance1} is the distribution of elements synthesized by the reference model 5p02822d40.16. 
The left panel displays the angle-averaged composition of the model, while the right panel shows the abundance distribution incorporating macroscopic mixing at 2000~\kms. This mixing enhances the overlap of the $^{56}$Ni distribution with Ca and, to a lesser extent, with other intermediate-mass elements.

\FloatBarrier

Table~\ref{tablemodellines} highlights the most significant contributors to the optical synthetic spectrum. It is important to note, however, that these features are heavily influenced by numerous blends.

{
\centering 
\small
\begin{longtable}[!h]{|crr|crr|crr|crr|}
\caption{List of Relevant Optical Forbidden Lines from the Mixed Model.
}
\label{tablemodellines}\\
\hline
Strength\tablefootmark{a} & $\lambda [\mu\mathrm{m}] $  &Ion  &   
Strength &  $\lambda [\mu\mathrm{m}] $  &Ion  &   
Strength &  $\lambda [\mu\mathrm{m}]$  &Ion  &   
Strength &  $\lambda [\mu\mathrm{m}] $  &Ion  \\  
\hline
\endfirsthead
\caption{continued.}\\
    \hline
Strength\tablefootmark{a} & $\lambda [\mu\mathrm{m}] $  &Ion  &   
Strength &  $\lambda [\mu\mathrm{m}] $  &Ion  &   
Strength &  $\lambda [\mu\mathrm{m}]$  &Ion  &   
Strength &  $\lambda [\mu\mathrm{m}] $  &Ion  \\  
\hline
\endhead
\hline
\endfoot
       &     0.4565 &    [\ion{Co}{ii}]  & 
   *   &     0.4567 &    [Ni I]   & 
   *   &      0.4569 &    [Co III] & 
   *   &      0.4571 &    [Co III] \\
       &     0.4575 &    [Ni II]  &
   *   &     0.4578 &    [Fe II]  & 
 **    &     0.4583 &    [Co III] & 
  **   &     0.4590 &    [S I]    \\
   *   &     0.4599 &    [Co I]   & 
       &     0.4608 &    [Fe III] &
       &     0.4613 &    [Co II]  & 
       &     0.4617 &    [Co I]   \\
   *   &     0.4623 &    [C I]    & 
 ***   &     0.4624 &    [Co II]  & 
  **   &     0.4628 &    [Co III] &
       &     0.4629 &    [C I]    \\
       &     0.4629 &    [Ni II]  & 
  **   &     0.4629 &    [Ni II]  & 
       &     0.4631 &    [Co III] &
 ***   &     0.4641 &    [Fe II]  \\
  **   &     0.4659 &    [Co II]  & 
 ***   &     0.4660 &    [Fe III] & 
  **   &     0.4666 &    [Fe II]  & 
  **   &     0.4668 &    [Fe III] \\
   *   &     0.4685 &    [Co I]   & 
   **  &       0.4703 &    [Fe III] & 
  **   &     0.4712 &    [Ni I]   & 
       &     0.4715 &    [Co III] \\
   **  &      0.4719 &    [Co III] &
 ***   &     0.4729 &    [Fe II]  & 
       &     0.4735 &    [Co II]  & 
  **   &       0.4735 &    [Fe III] \\
\end{longtable}
\tablefoot{\tablefoottext{a}{For each transition, the markers indicate the following levels of detectability: strong (\( *** \)), moderate (\( ** \)), weak (\( * \)), and scarcely detectable (\( \, \)), superimposed on the quasicontinuum created by a large number of lines. The relative strength (\( S \)) is determined by integrating the envelope, expressed as
$\int A_{ij} \times n_j \, dV$,
where \( n_j \) represents the particle density of the upper level. {The qualitative indicators for line strength are approximately more than 50\%, 20\%, 10\%, 5 \% and less than 5\% of the maximum photon emissivity of features relative to the wavelength range shown in Fig.~\ref{fig:theory}}. The rest wavelength is provided in micrometers (\( \mu\mathrm{m} \)), along with the identification of the corresponding ion. Most spectral features are heavily blended, except for the feature associated with [Fe II] at 1.644 \( \mu\mathrm{m} \), which remains distinguishable due to the Doppler shift smearing each transition by approximately 3\%. The transitions are listed in ascending order of wavelength from left to right.}}
}

\end{appendix}

\end{document}